\documentclass[usenatbib]{mn2e}

\usepackage{graphicx}
\usepackage{lscape}
\usepackage{rotating}
\usepackage{scalefnt}

\newif\ifAMStwofonts

\bibpunct{(}{)}{;}{a}{}{,}

\newcommand{\kms}{km~s$^{-1}$\,}
\newcommand{\kmsdeg}{km~s$^{-1}$~deg$^{-1}$\,}

\newcommand {\sm}{\rm\,M$_\odot$}
\newcommand {\sL}{\rm\,L$_\odot$}
\newcommand {\mgi} {Mg\,{\small I}\,}

\title[The DART survey of the 
Sextans dSph]{Study of the Sextans dwarf spheroidal galaxy from the DART Ca~II triplet 
survey\thanks{Based on FLAMES observations collected at the ESO, proposal 171.B-0588.}}
\author[G.~Battaglia et al.]{
G.~Battaglia,$^{1}$\thanks{Corresponding author. E-mail: gbattagl@eso.org}
E.~ Tolstoy,$^{2}$
A.~ Helmi,$^{2}$
M.~Irwin,$^{3}$
P.~Parisi,$^{4}$
V.~Hill,$^{5}$
P.~Jablonka$^{6}$
\\
$^{1}$European Organization for Astronomical Research in the Southern Hemisphere, K. Schwarzschild-Str. 2, 85748 Garching, Germany\\
$^{2}$Kapteyn Astronomical Institute, University of Groningen, P.O.Box 800, 9700 AV Groningen, the Netherlands \\
$^{3}$Institute of Astronomy, Madingley Road, Cambridge CB03 0HA, UK \\
$^{4}$INAF - Istituto di Astrofisica Spaziale e Fisica Cosmica di Bologna, via Gobetti 101, 40129 Bologna\\
$^{5}$Universit\'e de Nice Sophia-Antipolis, CNRS, Observatoire de la C\^{o}te d'Azur, Laboratoire Cassiop\'ee,  B.P. 4229, 06304 Nice cedex 4, France\\
$^{6}$Observatoire de Gen\`eve, Laboratoire d'Astrophysique de l'Ecole Polytechnique F\'ed\'erale de Lausanne (EPFL), \\ 
CH-1290 Sauverny, Switzerland\\
}

\begin{document}

\date{Accepted 2010 September 17.  Received 2010 September 6; in original form 2010 June 25}

\pagerange{\pageref{firstpage}--\pageref{lastpage}} \pubyear{2002}

\maketitle

\label{firstpage}

\begin{abstract}
We use VLT/FLAMES 
intermediate resolution (R$\sim$6500) spectra of individual red giant branch stars in the near-infrared Ca\,{\small II}\, triplet (CaT) region to investigate 
the wide-area metallicity properties and internal kinematics of the Sextans dwarf spheroidal galaxy (dSph). 
Our final sample consists of 174 probable members of Sextans with accurate line-of-sight velocities ($\pm$ 2 \kms) and CaT [Fe/H] measurements ($\pm$ 0.2 dex). 
We use the Mg~\,{\small I}\, line at 8806.8 \AA\, as an empirical 
discriminator for distinguishing between probable members of the dSph (giant stars) and probable Galactic contaminants (dwarf stars).

Sextans shows a similar chemo-dynamical behaviour to other Milky Way dSphs, with its  
central regions being more metal rich than the outer parts and with the more metal-rich stars displaying colder kinematics than the 
more metal-poor stars. 

Hints of a velocity gradient are found along the projected major axis and along an axis at P.A.$=191^{\circ}$, however a 
larger and more spatially extended sample may be necessary to pin down the amplitude and direction of this gradient.

We detect a cold kinematic substructure at the centre of Sextans, consistent with being the remnant 
of a disrupted very metal poor stellar cluster.

We derive the most extended line-of-sight velocity dispersion profile for Sextans, out to 
a projected radius of 1.6 deg. From Jeans modelling of the 
observed line-of-sight velocity dispersion profile we find that this is consistent with both a cored dark matter halo  
with large core radius and cuspy halo with low concentration. The mass within the last measured point 
is in the range 2-4$\times 10^8$ \sm, giving very large mass-to-light ratios, from 460 to 920 (M/L)$_{V,\odot}$.

\end{abstract}

\begin{keywords}
stars: abundances -- galaxies: kinematics and dynamics -- galaxies: dwarf -- Local Group 
-- Galaxies: individual: Sextans dSph -- dark matter
\end{keywords}

\section{Introduction}
The Sextans dwarf spheroidal galaxy (dSph) is a satellite of the Milky Way
(heliocentric distance of 86 kpc from \citealt{irwin1990} and \citealt{mateo1995}; 95.5 kpc from \citealt{lee2003}). 
Sextans was discovered relatively recently by automatic scanning of APM plates by \citet{irwin1990}. 
This is because the Sextans dSph (hereafter, Sextans) is one of the most diffuse and faint dSphs, 
with a central surface brightness $\Sigma_0 = 18.2 \pm 0.5$ mag arcmin$^{-2}$ 
and luminosity $L= 4.1 \pm 1.9 \times 10^5$ \sL \citep{IH1995}. 
This, together with its location on the sky \citep[$l= 243.5^{\circ}$, $b= +42.3^{\circ}$ from][]{mateo1998}, results in a considerable 
amount of contamination from Milky Way stars. 
This makes it particularly difficult to explore the 2D structure of Sextans and search for 
possible asymmetries or other features \citep{IH1995}. 

A further difficulty in the study of Sextans is given by its large extent 
on the sky ($r_{\rm core}=16.6 \pm 1.2$ arcmin; $r_{\rm tidal}= 160.0 \pm 50$ arcmin from \citealt{IH1995}), 
which is not well suited to the field of view of most current wide area imagers and spectrographs. 
Several imaging studies of the central region 
(at most the inner 40 arcmin $\times$ 40 arcmin) of Sextans 
\citep{mateo1991, mateo1995, bellazzini2001, harbeck2001, lee2003} 
have shown that the majority of its stars are ancient ($>$10 Gyr old), 
as evidenced by the presence of horizontal branch (HB) and RRLyrae stars. A  
significant population of stars have also been found on the main sequence (MS), above the 
oldest turn-off. It was suggested that these may be MS stars as young as 2 Gyr  \citep[e.g.][]{mateo1991}. However, 
deep Suprime-Cam $V$ and $I$ photometry at Subaru \citep{okamoto2008} shows that 
their spatial distribution is very similar to old ($>$10 Gyr) MS stars below the oldest 
turn-off, indicating that these 
are likely to be old blue stragglers (BS) stars ($>$10 Gyr) rather than young MS stars \citep[see also][]{lee2003}. 

In common with other dSphs, Sextans presents spatial variations of its stellar population mix, with 
the red horizontal branch (RHB) stars more centrally concentrated than the blue horizontal branch (BHB) stars 
\citep{bellazzini2001, harbeck2001, lee2003, kleyna2004}. By estimating the luminosity difference 
between the HB and the main-sequence turn-off out to the nominal tidal radius of Sextans, \citet{okamoto2008} estimate an age difference 
of at least 3 Gyr between 
the outer parts (with age $\sim$ 14 Gyr) and the central parts (with age $\sim$ 10 Gyr). The younger stars 
are more centrally concentrated than the older ones, consistent with the different distribution of RHB and BHB stars. 
The other populations,  red giant branch (RGB), BS and MS stars, 
have a spatial distribution intermediate 
to the one of RHB and BHB stars: this indicates that their spatial distribution can 
be thought of as a combination of the stellar distribution of RHB and BHB stars, i.e. RGB, BS and MS are 
likely to represent both components.

No determinations have yet been made of the large scale   
metallicity properties in Sextans. Spectroscopic estimates of the mean metallicity ([Fe/H]) of Sextans are available 
from Ca\,{\small II} triplet (CaT) observations 
of individual red giant branch (RGB) stars in the central regions ($<$20 arcmin) of the galaxy 
\citetext{\citealt{dacosta1991}, $<$[Fe/H]$> = -1.7 \pm 0.25$ dex, 6 stars; 
\citealt{suntzeff1993}, $-2.05 \pm 0.04$ dex, 43 stars}, which also revealed  
an internal [Fe/H] spread. 
High resolution spectroscopic observations of 5 RGB stars in Sextans by \citet{shetrone2001} 
confirmed both the average [Fe/H] found in the CaT studies, 
and the large range of observed metallicities ($\Delta$[Fe/H]$=1.4$ dex). The derivation of the 
mean [Fe/H] and spread in Sextans by \citet{geisler1996} from Washington photometry is also in 
agreement with the spectroscopic studies. Recently \citet{aoki2009} and Tafelmayer et al. (2010) 
follow-up at high spectral resolution 6 and 2 stars in Sextans, respectively, confirming the 
presence of stars with [Fe/H] down to $-3.1$.

Studies on the internal kinematics of Sextans have shown that, 
in common with the other dSphs, this galaxy exhibits a much larger velocity dispersion than what is 
expected from 
the gravitational contribution of its luminous matter, 
hinting at the presence of large amounts of dark matter 
\citep{suntzeff1993, hargreaves1994, kleyna2004, walker2006}. 
The latest estimate \citep{walker2007} based on a velocity dispersion profile from 504 members 
gives a mass of 5.4$\times 10^7$ \sm within the last measured point at $\sim$65 arcmin 
(equivalent to $\sim$4 core radii, $\sim$0.4 tidal radii), 
i.e. a mass-to-light (M/L) ratio of 130 (M/L)$_{\odot}$. \citet{strigari2007} used 
a sample of 294 members from \citet{walker2006} and estimated a mass of $0.9_{-0.3}^{+0.4} \times 10^7$ \sm 
within 0.6 kpc ($13_{-5.8}^{+11} \times 10^7$ \sm extrapolated to 4 kpc), which corresponds to 
an M/L$=260$ (M/L)$_{\odot}$. Although the mass estimates are not directly comparable as they refer to different distances 
from the Sextans centre, they all point to mass-to-light ratios so large as to require considerable 
amounts of dark matter, if the hypothesis of dynamical equilibrium holds.  These 
spectroscopic samples do not reach the nominal tidal radius and 
therefore the masses determined are likely to be lower limits. 

With its unusually large extent on the sky and large tidal-to-core radius ratio 
($r_{\rm tidal}/r_{\rm core} \sim 9.6$ as compared to ratios between 3 and 6 for most of the 
other classical dSphs, see Irwin \& Hatzidimitriou 1995), Sextans seems 
a natural candidate for tidal disruption. However, no clear signs of disruption such as 
tidal tails and S-shaped contours have been found in Sextans. This may 
however just be a consequence of the observational challenge of detecting such signs 
in a low surface brightness and heavily contaminated object like Sextans, or 
that such signs are not necessarily expected in tidally disrupted galaxies, as some N-body simulations show 
\citep[see e.g.][]{munoz2008}. 
A small velocity gradient has been detected approximately along the projected minor 
axis of this galaxy by \citet{walker2008}. However, given that no proper motions 
have been directly measured for Sextans, it is unclear whether this gradient is  
a sign of tidal disruption as predicted, for example, in the models of \citet{oh1995}, 
due to geometrical effects \citep{strigari2010} or an intrinsic gradient.

In this paper we use a spectroscopic sample of line-of-sight velocities and [Fe/H] measurements 
for 174 RGB stars probable members of Sextans, obtained at VLT/FLAMES, to analyse both the 
wide-field metallicity properties and the internal kinematics of this galaxy. Our sample 
extends out to 6 core radii (0.7 times the nominal tidal radius). The structure of the paper is the following. 
In Sect.~\ref{sec:obs} we describe the observations, the 
data reduction technique and the adopted metallicity scale. In Sect.~\ref{sec:mem} we deal 
with the issue of separating Galactic contaminants from Sextans members and 
we use the Mg~I line as an empirical discriminator between dwarf and giant stars. In Sect.~\ref{sec:results} 
we present the results from our survey regarding the wide-area metallicity and kinematic properties of Sextans as 
well as the presence of kinematic substructures. Finally, we perform a mass determination in Sect.~\ref{sec:mass} and 
present discussion and conclusions in Sects.~\ref{sec:disc} and \ref{sec:summary}.

\section{Observations and data reduction}\label{sec:obs}
The spectroscopic observations were carried out at VLT/FLAMES between 2003-2004, 
with the exception of one pointing which was re-observed in May 2008, during the 
commissioning of the upgraded GIRAFFE spectrograph CCD.

For the selection of our spectroscopic targets we used photometric data  
from INT/WFC covering approximately an area of 2 deg$\times$2 deg centred on the 
coordinates of the optical centre galaxy and ESO/WFI photometry 
for 12 pointings spread over the outer regions of the system. The 
imaging data were reduced following standard procedures, for details 
see references in \citet{battaglia2006}. In this work these photometric data are used exclusively to select spectroscopic targets. 

We selected targets classified as stellar in our photometry and 
with a position on the colour-magnitude diagram (CMD) consistent with an RGB star, but with a wide colour range to avoid 
biasing our sample in age or in metallicity.

We used VLT/FLAMES feeding the GIRAFFE spectrograph 
in Medusa mode, that allows the simultaneous allocation of   
132 fibres (including sky fibres) 
over a 25' diameter field of view \citep{Pasquini2002}. We used the GIRAFFE low resolution grating 
LR8 (resolving power R$\sim$ 6500), covering the wavelength range from 8206 \AA\,  to 9400 \AA. 
This allows the
measurement of equivalent widths (EW) from the near-infrared CaT lines at 8498, 8542, and 8662 \AA\, and also enables the derivation of
velocities accurate to a few km/s. The spectra used in this work 
are from 16 different fields in Sextans obtained with this set-up 
(see Tab.~\ref{tab:journal} for the observations journal). 
These data were reduced using the GIRAFFE pipeline 
(Geneva Observatory; Blecha et al. \citeyear{blecha2003}). 

The sky-subtraction and 
extraction of the velocities and EWs of CaT lines 
were carried out using our own software 
developed by M.Irwin. For the details see \citet{battaglia2008a}. 
Here we just note that we estimate the CaT EW in two ways. The
first consists in simply summing the flux contained in a region 15\AA\ wide 
centred on each CaT line (hereafter, integral fit).  To
derive the second estimate we fit individual unconstrained Gaussian
functions to each CaT line over the same wavelength region (hereafter, Gaussian fit).  
The combined EWs for CaT lines \#2 and \#3 ($\lambda_{\rm 8542}$,
$\lambda_{\rm 8662}$) for both the integral and Gaussian fits
are then compared and used to compute an overall correction to the
Gaussian fit.  This is necessary since the observed CaT lines have non-Gaussian wings 
which are progressively more visible as the EW increases. 

Beside the EW of the CaT lines, we also measure the EW of a \mgi line at 8806.8 \AA\,. The 
EW of this line is estimated using the values from the integrated fit, 
summing the flux contained in a region 6\AA\ wide 
centred on the \mgi line. The EW of this line will 
be used as a discriminant between Sextans stars and Milky Way contaminants as described in Sect.~\ref{sec:mgI}.
 
We observed a total of 1294 targets, which include 103 stars with 
double measurements, 70 with 3 measurements, 5 with 4 measurements. 
The number of distinct targets is 1036. 

As described in our previous studies using the same instrument and set-up we 
ensured a high reliability for the data used in our analysis 
by excluding those stars with S/N per \AA\ $<$ 10 and 
velocity errors larger than 5 \kms 
\citep[e.g.][]{battaglia2008a}. In order to ensure the reliability of the 
EW measurements we also required that the summed EW of the two strongest CaT lines 
should not differ by more than 2 \AA\, between the determination from the 
integral fit and the Gaussian fit. 

\begin{figure*}
\begin{center}
\includegraphics[width=70mm]{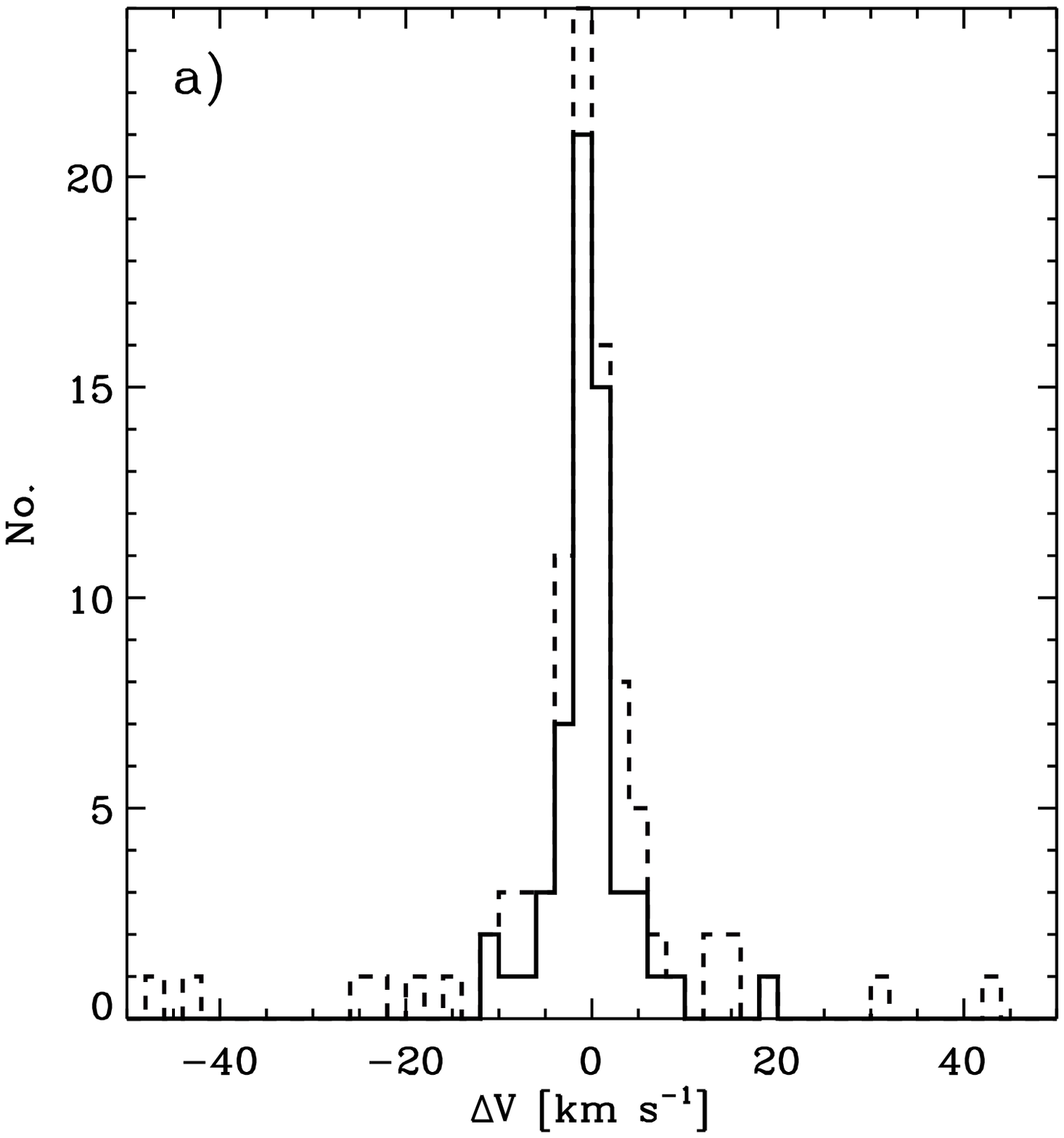}
\includegraphics[width=70mm]{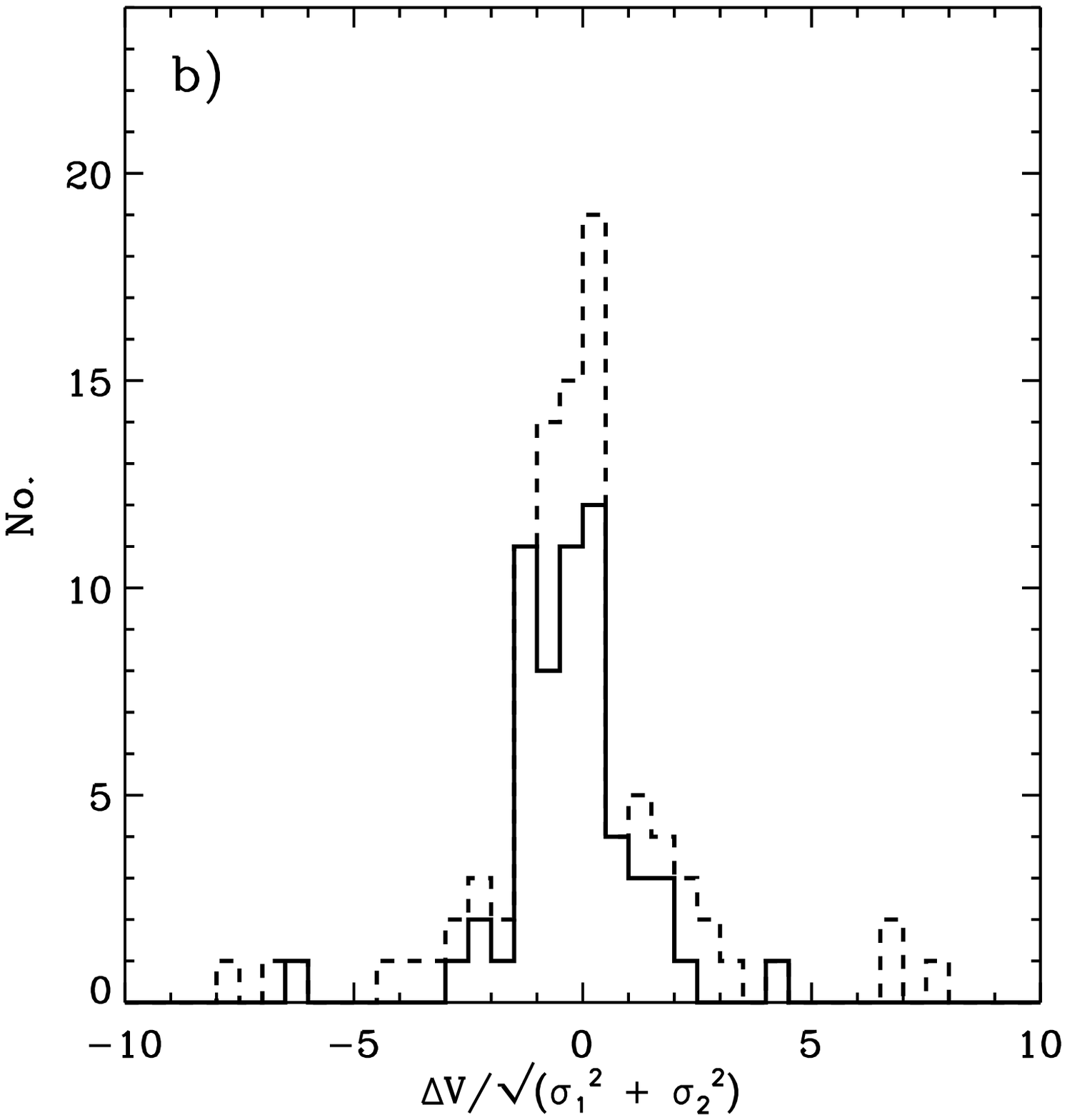}
\caption{Comparison between velocity measurements for stars with double measurements in the 
Sextans dSph. Panel a) Distribution 
of velocity differences for all the stars (dashed line), 
and for the stars with S/N per \AA\,$\ge$ 10 and 
estimated error in velocity $\le$ 5 \kms for each measurement (solid line). 
The weighted mean velocity, 
$rms$ dispersion and scaled 
median absolute deviation from the median (where the scaled median absolute deviation from the median (MAD) 
is 1.48$\times$MAD $\equiv$ a robust $rms$ e.g.  
Hoaglin et al. \citeyear{hoaglin1983}) 
are: $0.9 \pm 3.0$ \kms, $29.0 \pm 2.2$ \kms, $4.3$ \kms (dashed line) and $-0.6 \pm 0.5$ \kms, $2.5\pm 0.5$ \kms, 
$2.0$ \kms (solid line). Panel b) As above but now the velocity difference is normalised by the predicted error. 
With these S/N and velocity error cuts 
the measured error in the velocity distribution is close to the expected unit variance Gaussian
(1.48$\times$MAD = 0.98).
}
\label{fig:double_vel}
\end{center}
\end{figure*}

\begin{figure*}
\begin{center}
\includegraphics[width=70mm]{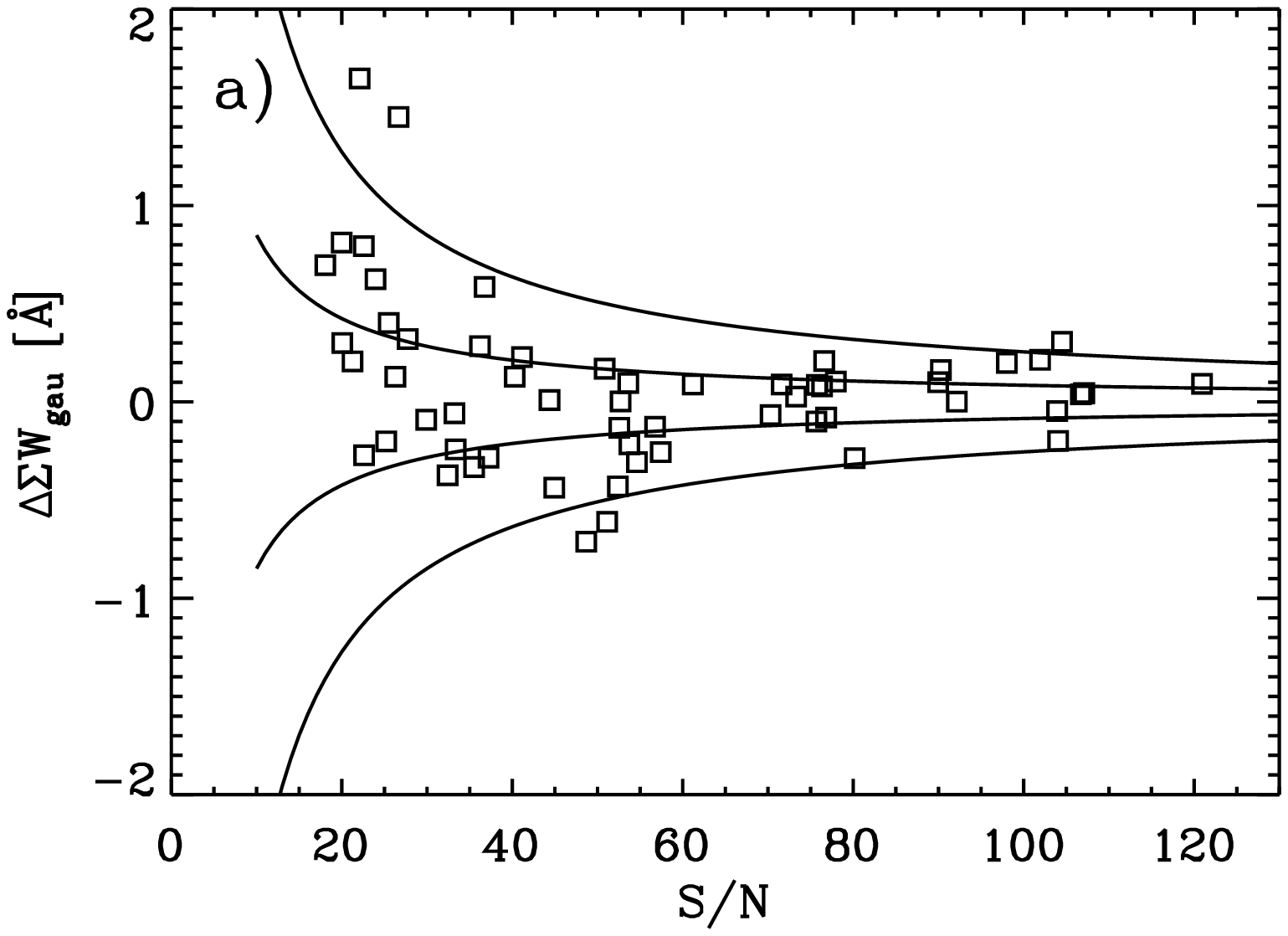}
\includegraphics[width=70mm]{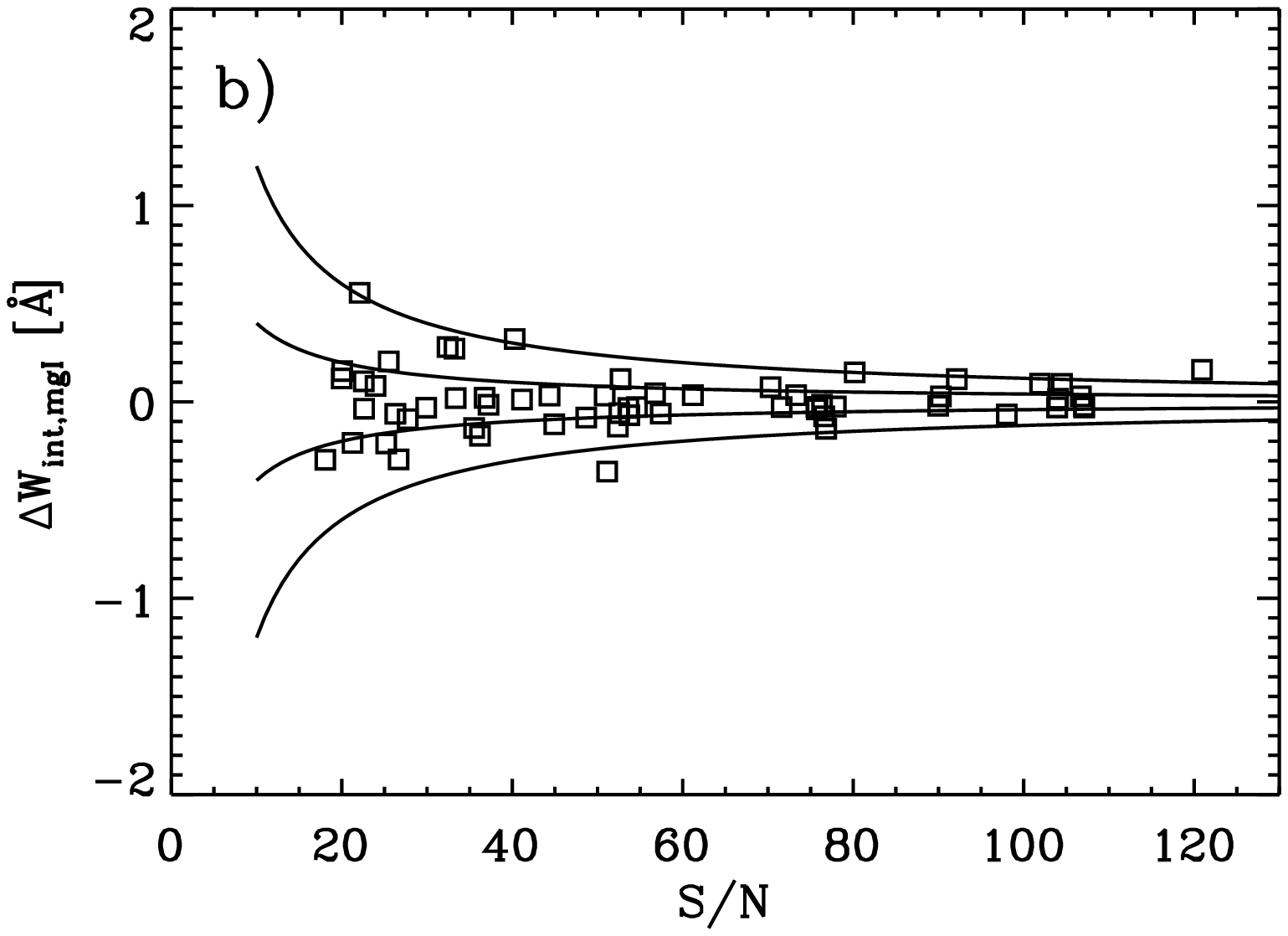}
\caption{
Comparison between EW measurements for stars with double measurements in the 
Sextans dSph. Distribution 
of summed CaT EW (EW$_2$+EW$_3$)  differences from Gaussian fit (left) 
and \mgi EW differences from integrated flux (right) as a function of S/N 
for the stars with S/N per \AA\,$\ge$ 10 and 
estimated error in velocity $\le$ 5 \kms for each measurement. Assuming a 
similar S/N for the individual measurements of  each star, 
the error in the summed CaT EW difference is $\sqrt{2} \times \sigma_{\Sigma W} = \sqrt{2} \times 6 / (S/N)$ 
while the error in the \mgi EW difference is $\sqrt{2} \times \sigma_{\Sigma W} = \sqrt{2} \times 2.8 / (S/N)$ . 
The solid lines indicate the 1 and 3 $\sigma$ 
region for the error in the difference of summed EW. The weighted mean summed CaT EW difference, 
$rms$ dispersion in the difference and scaled MAD are $0.03 \pm 0.05$ \AA\,, $0.22\pm 0.06$ \AA\,, 
$0.25$ \AA\,.  The weighted mean in the \mgi EW difference, 
$rms$ dispersion and scaled median absolute deviation from the median 
 are $0.01 \pm 0.02$ \AA\,, $0.09\pm 0.03$ \AA\,, 
$0.08$ \AA\,. 
} \label{fig:double_ew}
\end{center}
\end{figure*}

Figure~\ref{fig:double_vel} shows the distribution of differences 
in velocity for the stars that have double measurements (103 objects), both without 
any selection criteria and adopting the above selection criteria for each measurement (59 objects). 
Figure~\ref{fig:double_ew} 
refers to the distribution of differences in summed EW from Gaussian fit of CaT lines and integrated flux for the \mgi line. 
The determination of the summed CaT EW from the integral and Gaussian estimators yield quite similar distributions, 
and therefore, for consistency with our previous work, hereafter we use the summed CaT EW derived from the 
Gaussian estimator. For the EW of the \mgi line the integral estimator yields slightly less noisy measurements and 
therefore we will use this for the \mgi line. The error in the summed CaT EW for the 
individual measurements is given by $\sigma_{\rm \Sigma W}=$6/(S/N), while 
the error in the \mgi EW is well represented by 2.8/(S/N).

In the following, when calculating the velocity, summed CaT EW and all the other properties of 
the stars with repeated measurements, we use only the individual measurements which meet our S/N and velocity errors criteria 
(before combining them).
 
The final sample was carefully checked to weed out any spurious objects 
(e.g. broken fibres, background galaxies, foreground stars, etc.). We removed 
4 objects because they were observed with broken fibres; we found 3 background galaxies and removed 9 objects because the 
continuum shape or the presence of very broad absorption line was not consistent 
with what is expected for RGB stars. 

Our final sample of acceptable measurements for velocities and CaT EWs consists of 789 stars. 
We provide a detailed comparison between this data-set and the one from \citet{walker2009} in the 
Appendix.

\subsection{Metallicity scale}
We use the CaT EWs to derive metallicities ([Fe/H]) for the target stars.  

The near-infrared CaT lines have been extensively used in the literature as 
empirical estimators of the [Fe/H] abundance of individual RGB stars 
observed at intermediate spectral resolution. 

This method has been empirically calibrated for stellar 
clusters and proven reliable over the range $-2.1 \la $[Fe/H]$\la -0.2$ and $2.5 \la $ age [Gyr] $\la 13$ 
\citep[e.g.][]{rutledge1997b, cole2004} on the \citet{carretta1997} metallicity scale. 
\citet{battaglia2008a} tested the validity of the method for composite stellar populations, 
observing the same 129 individual stars in the Sculptor and Fornax dSphs both at intermediate and high resolution,  
and showed that the CaT EW- [Fe/H] relation can be applied to composite 
stellar populations over the explored range $-2.5 \la $[Fe/H]$\la -0.5$. 

\citet{starkenburg2010} carried out a synthetic spectral analysis of the CaT lines down to [Fe/H]$= -4$ in order to understand their theoretical 
behaviour  as a function of physical parameters such as metallicity, gravity, 
effective temperature and possible biases in the low metallicity range. The authors 
find that the a simple linear relation no longer holds 
approximately below [Fe/H]$\la -2.5$. This behaviour is mostly a reflection of the change 
of the CaT lines from wing-dominated at high metallicity to core-dominated at low metallicity. 

The authors provide a revised relation, which 
holds over the much larger range $-4.0 \la $[Fe/H]$\la -0.5$:
\begin{eqnarray}
{{\rm [Fe/H]}= -2.87  + 0.195 \times (V - V_{\rm HB}) + 0.458 \times \Sigma W +{}}
                        \nonumber\\
{}  - 0.913 \times {\Sigma W}^{-1.5} + 0.0155  \times {\Sigma W} \times (V - V_{\rm HB})
\label{eq:catfeh_else} 
\end{eqnarray}
In the above $\Sigma W$ is the summed EW of the 
two strongest CaT lines, i.e. $\Sigma W = EW_{2} + EW_{3}$, $V$ is the apparent magnitude in V-band of the 
star and $V_{\rm HB}$ is the apparent magnitude in V-band of the horizontal branch of the galaxy. 
For Sextans we use $V_{\rm HB} = 20.35$ from \citet{IH1995}, which is also consistent with our photometry.

Using our sample of 789 stars 
along the line-of-sight to Sextans, we find that the [Fe/H] values 
derived using Eq.~(16) in \citet{battaglia2008a} and Eq.~(\ref{eq:catfeh_else}) 
compare very well down to [Fe/H]$=-2$; at lower metallicities, Eq.~(16) from \citet{battaglia2008a} provides on average a larger 
[Fe/H] value than the revised calibration given by Eq.~(\ref{eq:catfeh_else}), e.g. 
a [Fe/H]$\sim -2.4$ with the old calibration corresponds to approximately [Fe/H]$\sim -2.6$ 
with the revised one and [Fe/H]$\sim -2.6$ to about [Fe/H]$\sim -3.0$.

In the following we will use Eq.~(\ref{eq:catfeh_else}) for our CaT [Fe/H] determinations. 
In order to derive the errors in [Fe/H], we follow the approach of \citet{starkenburg2010}, 
i.e. we consider that the error due to the photometry is negligible with respect to the error in 
the determination of the CaT EW and we input in Eq.~(\ref{eq:catfeh_else}) the values of 
$\Sigma W - 1 \sigma_{\rm \Sigma W}$ and $\Sigma W + 1 \sigma_{\rm \Sigma W}$ and derive the corresponding 
$1 \sigma$ lower and upper limit in [Fe/H]. This reflects the fact that the 
relation is not linear and therefore symmetric errors in the EW determination can translate into 
asymmetric errors in [Fe/H].

\section{Membership} \label{sec:mem}
Given the low surface brightness of Sextans and its location on the sky, the RGB locus of Sextans is heavily contaminated by Milky Way stars 
which are located along the line-of-sight to Sextans. In order to study the metallicity and kinematic properties of 
Sextans it is important that we first weed out these Galactic interlopers 
from our sample. These interlopers will be mostly foreground MW stars. 
Given the Galactic coordinates of Sextans, the Besan\c con model \citep{robin2003} predicts 
that the majority of interlopers along the line-of-sight to Sextans 
 is made of dwarf stars, mostly from the thin and thick disk,  and by smaller fraction by sub-giants and giant stars.

Ideally 
one would like a method allowing a direct discrimination between members of the dSph 
(which have been selected to be RGB stars) from non-members 
(mostly dwarf stars) on the basis of, for example, spectral features 
indicating physical characteristics, such as 
in this case gravity. However, no such spectral features have been 
identified in previous studies using a similar spectral 
range and set-up. The common approach is therefore statistical. 

\begin{figure*}
\begin{center}
\includegraphics[width=90mm]{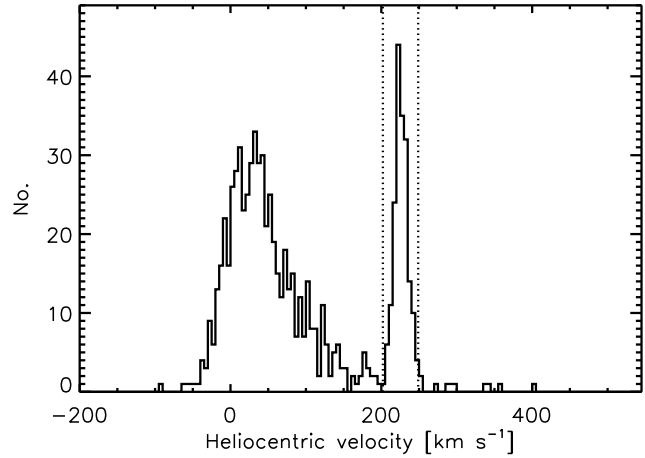}
\caption{Distribution of heliocentric velocities for the stars observed with VLT/FLAMES in the direction of 
Sextans and that met our selection criteria. The dotted lines indicate the region used for the 3-$\sigma$ membership 
selection.}
\label{fig:histovel}
\end{center}
\end{figure*}

The simplest way of discerning between members and non-members is to consider 
the line-of-sight velocity distribution of the observed targets (see Figure~\ref{fig:histovel} for the 
distribution of heliocentric velocities for 
our VLT/FLAMES targets) and 
iteratively apply a k-$\sigma$ cut around the systemic velocity of the galaxy, 
where $\sigma$ is the internal line-of-sight velocity dispersion of the galaxy. The k-$\sigma$ cut 
in common use is the 3$\sigma$ cut, justified by the close to Gaussian line-of-sight velocity distribution 
observed for dSphs. 

We first follow this approach and then refine our selection using further criteria.

\subsection{3-$\sigma$ clipping} \label{sec:3sigma}
We first identify the peak of the velocity distribution and derive the weighted average velocity and 
dispersion for the stars with a velocity within 4$\sigma$ of the peak, using as a first guess 
a broad dispersion of $\sigma= 15$ \kms. We then repeat the procedure restricting 
the range of selection to 3$\sigma$. We then apply to the distribution of line-of-sight velocities 
for the stars within the 3$\sigma$ range the maximum likelihood 
method outlined in \citet{hargreaves1994}. This results in a systemic velocity  
$v_{\rm sys, hel}= 226.3 \pm 0.6$ \kms and a global dispersion $\sigma_{\rm hel} = 8.8 \pm 0.4$ \kms 
for 182 probable members. The systemic velocity derived is in very good agreement with the 
value of $v_{\rm sys, hel}= 225.8 \pm 0.5$ \kms from \citet{walker2006}; for the global 
dispersion they quote values from 3 sample sizes: 
N= 276 $\sigma_{\rm hel} = 7.1 \pm 0.3$ \kms; N=294 $\sigma_{\rm hel} = 8.9 \pm 0.4$ \kms; 
N=303 $\sigma_{\rm hel} = 10.3 \pm 0.5$ \kms. Our determination is in very good 
agreement with their determination from the N=294 sample. We find a good agreement also with the 
values derived by \citet{hargreaves1994} ($v_{\rm sys, hel}= 224.4 \pm 1.6$ \kms, $\sigma_{\rm hel} = 7.0_{-1.0}^{+1.3}$\kms 
from 21 members), and consistent values also to the study of \citet{suntzeff1993} ($v_{\rm sys, hel}= 227.9 \pm 1.8$ \kms; 
$\sigma_{\rm hel} = 6.2 \pm 0.9$\kms based on 33 stars). 

Given the large extent on the sky of Sextans, 
the Solar motion and Local Standard of Rest (LSR) will contribute 
a component along the line-of-sight to the object, which may introduce spurious 
velocity gradients and affect the derivation of quantities such as internal rotation for example.
 In order to remove this effect 
we transform the heliocentric line-of-sight velocities 
into line-of-sight velocities in a frame at rest with respect to the Galactic centre 
($v_{\rm GSR}$ where GSR stands for Galactic standard of rest). 
For this we use the formula in \citet{binney1987}, using a 
LSR velocity $v_{\rm LSR}= 220$ \kms at the Solar radius ($R_{\odot}=$ 8kpc) and a Solar motion 
of ($U, V, W$)$=$(10, 5.25, 7.17) \kms \citep{dehnen1998}, where $U$ is radially inward, $V$ 
positive in the direction of Galactic rotation and $W$ towards the North Galactic Pole. 

We repeat the determination of the systemic velocity and global dispersion using the 
GSR velocities, and we obtain $v_{\rm sys, GSR}=78.6 \pm 0.7$ \kms and $\sigma_{\rm GSR}=8.8 \pm 0.5$ \kms 
for 182 probable members. Here after we will use the GSR velocities, 
except when comparing the individual 
line-of-sight velocities from this study those from other studies to ease the comparison.

\subsection{Use of the \mgi line at 8806.8 \AA\, as a dwarf/giant discriminator} \label{sec:mgI}
\begin{figure*}
\begin{center}
\includegraphics[width=90mm]{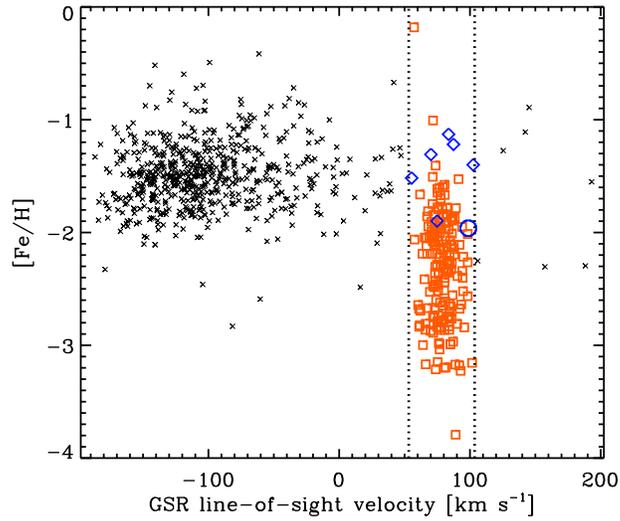}
\caption{[Fe/H] versus GSR line-of-sight 
velocity for the observed VLT/FLAMES targets (crosses and squares show the stars outside and within the 
3$\sigma$ membership region, respectively; the vertical dotted lines show the velocity membership region). Note that the stars found outside the 3$\sigma$ membership region, 
which are probable Galactic contaminants, tend to be preferentially found in the range $-2.0 \la {\rm [Fe/H]} \la -0.5$. 
This means that Sextans stars at ${\rm [Fe/H]} \la -2.0$ are less likely to be contaminated by Galactic stars whose velocity 
falls in the 3$\sigma$ membership selection region. The diamonds indicate stars with velocity consistent with membership but with 
\mgi EW too large for giant stars (see text).The circle indicates one object with velocity and \mgi EW consistent 
with membership but with unusual position on the CMD (see text).}
\label{fig:fevel}
\end{center}
\end{figure*}

Figure~\ref{fig:fevel} shows the distribution of [Fe/H] measurements for targets 
along the line-of-sight to Sextans against their  
line-of-sight velocity in the GSR system\footnote{Note that the CaT relation relies on the assumption that all the stars can be considered 
at the same distance. While this is a reasonable assumption for the stars belonging to the 
Sextans dSph, this is most likely not the case for the foreground/background MW stars. Therefore, 
the [Fe/H] values for the stars which are probable non-members cannot be considered as meaningful.}. 
Around the systemic velocity of Sextans $\pm$ 3$\sigma_{\rm GSR}$ one can distinguish the stars that are probable 
members of the galaxy, from those likely to be foreground/background MW stars. 
From Fig.~\ref{fig:fevel} it is clear that a simple kinematic selection like the 3-$\sigma$ clipping is able to 
remove the majority of the MW contaminants but that a small fraction may still be present 
in the 3-$\sigma$ region of membership selection. The fraction of contaminants appears to 
change with the [Fe/H] value derived from the CaT lines: at [Fe/H]$\la -2.0$ no considerable 
contamination is expected, while at larger metallicities Sextans stars are more likely 
to be contaminated by Galactic stars whose velocity 
falls in the 3$\sigma$ membership selection region. 

Here we attempt a new empirical approach to refine our membership determination, using the \mgi line at 8806.8\AA\, as a further possible discriminator 
between MW and Sextans members. 

\begin{figure*}
\begin{center}
\includegraphics[width=150mm]{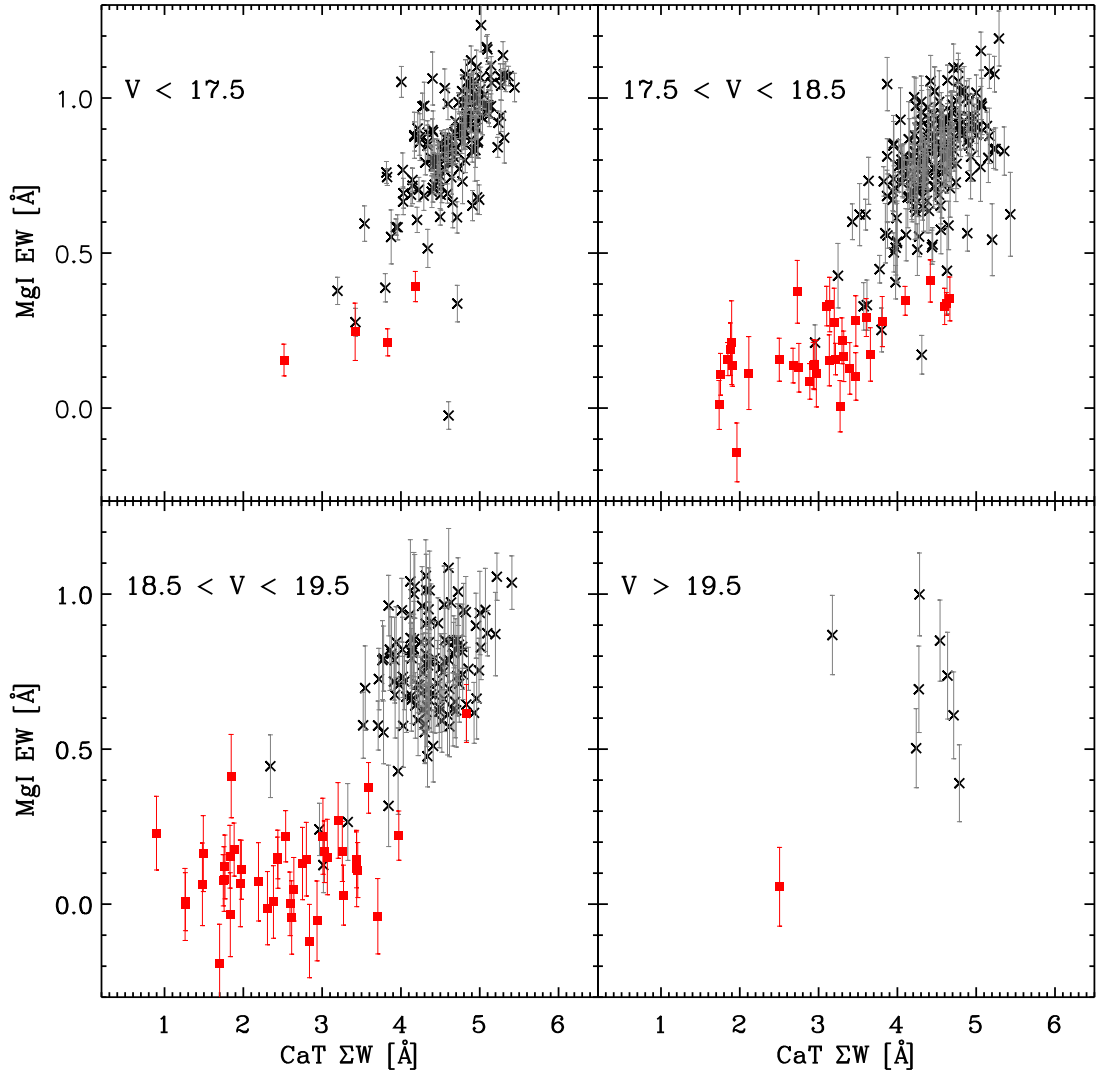}
\caption{\mgi EW vs CaT $\Sigma W$ in bins of V magnitude. The range of magnitudes are indicated on the top left corner of the panels. 
The filled red squares show the stars which are highly probable Sextans members (i.e. with velocities within 
$\pm 2 \sigma$ from the systemic velocity of Sextans) and the 
open black squares the stars which are highly likely to be MW contaminants (i.e. with velocities 
at least $4 \sigma$ away from the systemic velocity of Sextans).}
\label{fig:mgI}
\end{center}
\end{figure*}

We first select 
the stars that have a high probability of membership to Sextans by restricting the 
velocity range of selection to 2$\sigma_{\rm GSR}$ just for the purpose of this analysis; 
the highly likely non-members are chosen to be the ones with GSR  
velocity more than 4$\sigma_{\rm GSR}$ away from Sextans systemic velocity. 
Given the small EW of the \mgi line, in general less than 1 \AA\,, only spectra 
with S/N/\AA\, $>$ 20 are used to explore possible trends without being affected considerably by the 
noise. In order to explore the behaviour of the \mgi line with gravity in an empirical way, in 
Fig.~\ref{fig:mgI} we show the behaviour of the EW of the \mgi line versus the CaT 
$\Sigma W$ in bins of V-magnitude, for both high likely members and non-members. 

As it is visible in Fig.~\ref{fig:mgI} the loci of \mgi EW are quite distinct for members and non-members: members 
cluster at EW between $\sim$0 and 0.2 \AA\,, while non-members occupy preferably 
the region of \mgi EW between 0.5 and 0.9 \AA\,. Some members have negative \mgi EW: this is 
just a reflection of the lower S/N of those spectra; considering the error-bars these are 
consistent with \mgi EW close to zero, therefore with the line being hidden in the noise. 

It is clear that in general the 
non-members display a larger \mgi EW with respect to members. Preliminary use 
of synthetic stellar models suggest that this difference may not be driven exclusively by 
gravity but also of the different metallicity of the dSph star and MW stars (
with the former being below [Fe/H]$<-1$ and the latter peaking around $-0.5$ to $\sim$0). 

From the figure it appears that the trend of \mgi EW versus CaT 
$\Sigma W$ changes slightly among the bins in V-magnitude, which 
on a first approximation indicates a dependence on the gravity. However, 
a safe selection is to 
consider as members the stars with \mgi EW $<$ 0.5 \AA\,. Hereafter we 
add this constrain to our selection criteria for probable Sextans stars. We have checked 
that this selection is valid also when exploring the behaviour of the \mgi EW versus CaT 
$\Sigma W$ in bins of V$-$I, as a proxy for effective temperature. 

Applying this selection, 6 stars 
classified as members from the 3$\sigma$ clipping appear to be contaminants 
(these are indicated as diamonds in Fig.~\ref{fig:fevel}). 

We will improve on the simple empirical approach presented in this section using synthetic 
spectral libraries in order to assess the applicability range of this method.

\subsection{Location on the CMD}
\begin{figure*}
\begin{center}
\includegraphics[width=0.5\textwidth]{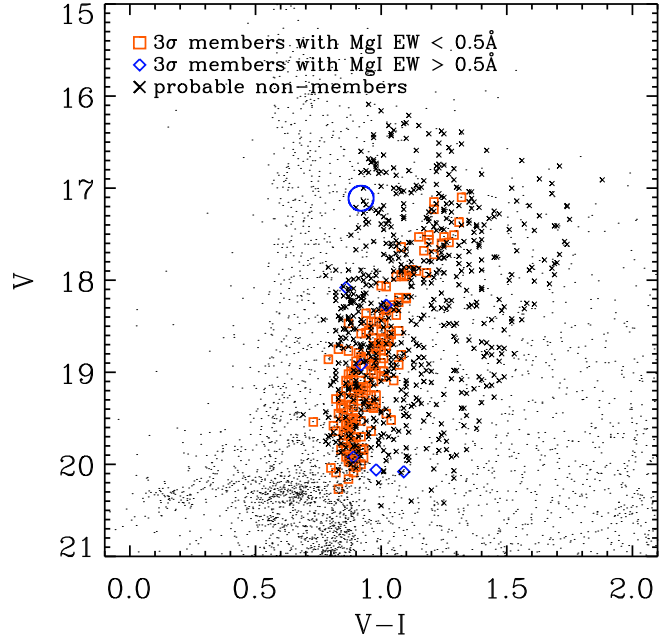}
\caption{Colour-magnitude diagram of stellar objects for Sextans as obtained from INT/WFC and ESO/WFI 
imaging data (dots), with overlaid the targets of our VLT/FLAMES spectroscopic observations 
which passed our quality criteria. The squares indicate the probable Sextans members, the diamonds the stars 
that would have been classified as members from the 3$\sigma$ clipping but that have a too large \mgi EW ($>$ 0.5 \AA\,), 
the crosses indicate the stars with velocities outside the 3$\sigma$ range of selection. The star encircled has a discrepant 
location on the CMD with respect to the other Sextans members of all metallicities, and it is the outermost star 
in our sample with a projected radius of more than 2 deg from Sextans centre. We believe that this star is likely to be 
a contaminant as well.}
\label{fig:cmd_spe}
\end{center}
\end{figure*}

Our VLT/FLAMES targets have been chosen to have magnitudes and colours broadly consistent with 
that expected for RGB stars. We can now apply the membership criteria 
of the previous two sections to redefine the RGB locus occupied by 
probable Sextans members. Figure~\ref{fig:cmd_spe} shows the location on the CMD 
of the probable members and probable non-members, selected using the criteria of Sects.~\ref{sec:3sigma} and \ref{sec:mgI}. 
All the probable members are closely located along the RGB ridge. The only evident 
outlier is found at V-I$\sim$ 0.9 and V$\sim$17.2 (see Fig.~\ref{fig:cmd_spe}). 
This star is located at a projected distance $R\sim$2.2 deg. 
Given its metallicity of [Fe/H]$=-1.9$, its location on the CMD appears inconsistent 
with stellar evolution models; we therefore exclude this star from subsequent analysis. 

\subsection{Final sample of probable members} \label{sec:finalmem}
\begin{figure*}
\begin{center}
\includegraphics[width=0.5\textwidth]{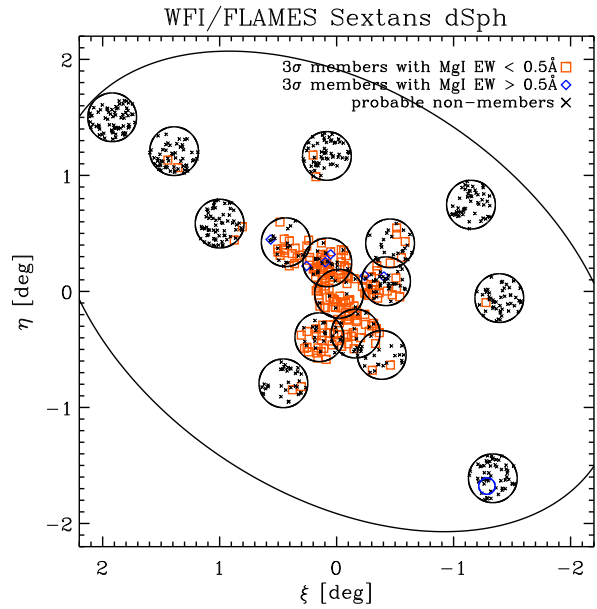}
\caption{Location of the observed VLT/FLAMES fields (solid circles) and targets in Sextans 
(squares: probable members; diamonds: stars with velocities within the 3$\sigma$ range of selection but 
with too large \mgi EW for giant stars; crosses: non-members with velocities outside the 3$\sigma$ range of selection). 
The small blue circle indicate the star with peculiar location on the CMD. The ellipse shows the nominal tidal radius 
\citep[value from][]{IH1995}.
}
\label{fig:fov}
\end{center}
\end{figure*}

By adopting all the membership criteria described above, i.e. 3$\sigma$ velocity clipping, \mgi EW and 
location on the CMD, the final sample of probable members is 174 stars. We have also 
excluded one star having an error in [Fe/H] $>$ 1 dex.
Using this sample the systemic velocity and global dispersion in the GSR system are 
$v_{\rm sys, GSR}=78.4 \pm 0.6$ \kms and $\sigma_{\rm GSR}=8.4 \pm 0.4$ \kms, while in the 
heliocentric system they are $v_{\rm sys, hel}= 226.0\pm0.6$ \kms and a global dispersion 
$\sigma_{\rm hel} = 8.4 \pm 0.4$ \kms, all in good agreement with previous determinations. 
 Hereafter we will use this sample for the analysis 
of the properties of Sextans. The location on the field-of-view of the cleaned up sample 
and the contaminants is shown in Fig.~\ref{fig:fov}. From this plot the problems in 
collecting large samples of individual stars in an object like Sextans are clear: 
it is very extended, therefore many pointings of wide-field area spectrographs like 
FLAMES are needed, and because of its very low surface brightness, the outermost 
3 pointings did not yield any probable members.

\section{Results} \label{sec:results}

\subsection{Metallicity properties} \label{sec:met}
Metallicity measurements of large numbers of individual stars are important since 
they can shed light on the processes that drive galactic chemical enrichment and by measuring the metal abundance  
in stars over a wide age range and across the galaxy one can recover how the metal content built up in time 
and how the enrichment proceeded throughout the galaxy.

\begin{figure*}
\begin{center}
\includegraphics[width=\textwidth]{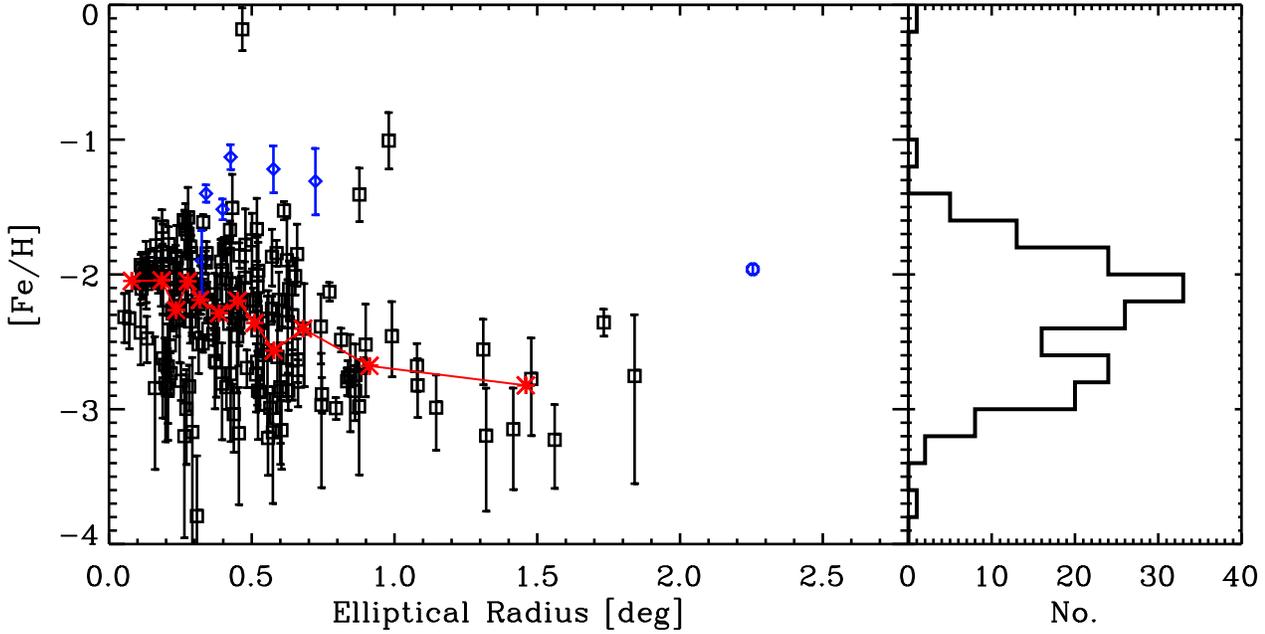}
\caption{Left: Metallicity distribution as function of elliptical radius for the probable members of Sextans 
(squares with error-bars). The diamonds with error-bars show those stars whose velocities fall within the 
3$\sigma$ range of membership, but that are likely non-members based on their large \mgi EW ($>$ 0.5 \AA\,). 
The small blue circle indicate the star with peculiar location on the CMD. Since the stars 
represented with the diamonds and the blue circle are likely non-members they will not 
be considered when deriving properties relative to Sextans. The red asterisks connected by a solid line 
represent a running median over 15 stars (except for the last point, which is over 9 stars. 
Right: Metallicity distribution for 
Sextans members (from the squares in right-handside panel).}
\label{fig:met}
\end{center}
\end{figure*}

Figure~\ref{fig:met} shows the metallicity distribution of probable members of Sextans as 
derived from our VLT/FLAMES data 
(left: as a function of elliptical radius; right: overall distribution). The stars 
in Sextans cover a wide range of [Fe/H] values, with the majority 
covering values from about $-3.2$ to $-1.4$, but reaching down to $-3.8$ and up to $-0.1$;  
the average value of the distribution is [Fe/H]$_{\rm avg} = -1.9$, with a scatter of 
0.6 dex, and the median value is [Fe/H]$_{\rm med}= -2.3$. 

The left panel of Fig.~\ref{fig:met} clearly shows that the metallicity properties of this 
galaxy change with the projected radius from the centre: the inner parts of the galaxy are more 
metal rich than the outer parts. This kind of behaviour is not unique to Sextans but has been detected also in 
other Milky Way satellites, on the basis of CaT data. However, 
here in Sextans the behaviour is more enhanced than in the other systems, with a 
step like behaviour of the metallicity distribution going from the region 
at R$<$0.8 deg to the region at R$>$0.8 deg: while the inner parts (R$<$0.8 deg) show a broad distribution, 
covering the whole range of [Fe/H] values, 
the stars in the outer parts (R$>$0.8 deg) appear to have all 
[Fe/H] values $< -2.2$ (except for 2 stars out of 19). The current data appear to suggest a [Fe/H] threshold ([Fe/H]$=-2.2$)
in the outer parts, but more data would be required to confirm that this result is not  
due to small number statistics. 
 
It is likely that 
the two ``metal-rich'' stars at R$>$0.8 deg are contaminants: their [Fe/H] value is very discrepant 
with respect to the rest of the distribution, 
with metallicities of about 1 dex above the largest [Fe/H] 
of the rest of the stars. These stars may be 
RGB stars belonging to the stellar halo of our Galaxy. Similarly, 
at R$\sim$ 0.5 deg, one star is present with 
[Fe/H]$\sim -0.1$, more than 1 dex larger than the maximum [Fe/H] displayed by all the other 
probable Sextans members anywhere in the galaxy. Even though it is possible that these 
3 stars may belong to the stellar halo of the Milky Way, these objects full-fill all our membership criteria and 
we have no means to distinguish 
them in a direct way from genuine Sextans members, therefore we will keep them in the analysis.

\begin{figure*}
\begin{center}
\includegraphics[width=0.6\textwidth]{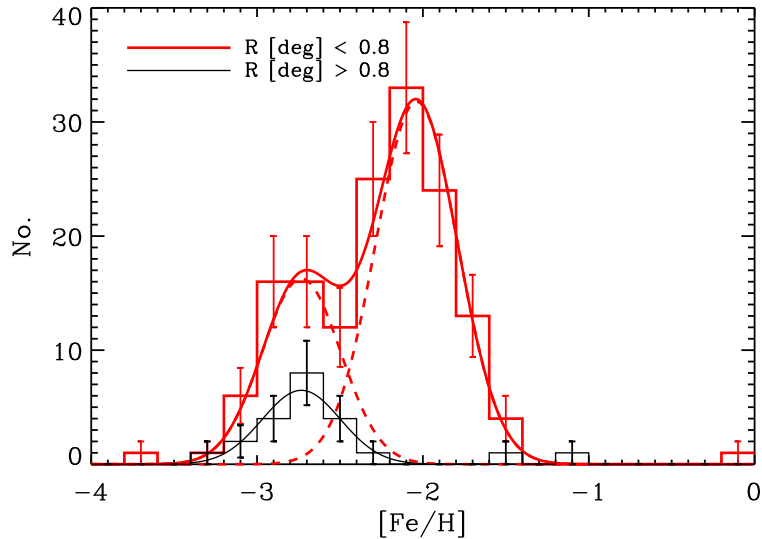}
\caption{Metallicity distribution for Sextans probable members at projected radii R$<$0.8 deg (thick red histogram) and 
at R$>$0.8 deg (thin black histogram). The solid black thin line shows the Gaussian fit to the distribution at 
R$>$0.8 deg; the thick solid red line shows the best fit to the distribution at 
R$<$0.8 deg for a sum of two Gaussians; the position of the central [Fe/H] and dispersion of one of the two Gaussian 
is kept constant to the values found for the distribution at R$>$0.8 deg. The dashed lines show the two Gaussians 
separately. The error-bars are Poissonian. }
\label{fig:histo_met}
\end{center}
\end{figure*}

As shown in Fig.~\ref{fig:histo_met} the metallicity distribution of R$>$0.8 deg 
is well approximated by a Gaussian with best-fitting peak position at [Fe/H]$=-2.73 \pm 0.06$ and 
dispersion $0.24 \pm 0.05$ dex (reduced $\chi^2 = 0.6$). The same figure shows that the region at R$<$0.8 deg is 
well represented by a sum of two Gaussians: one having the same 
shape parameters -modulo the amplitude - 
as in the outer parts, and the other Gaussian with best-fitting peak position at [Fe/H]$=-2.04 \pm 0.03$ and 
dispersion $0.25 \pm 0.03$ dex (reduced $\chi^2 = 0.7$). At R$<$0.8 deg these two Gaussians do overlap 
over a metallicity range and at [Fe/H]$\sim -2.3$ the contribution of the stars belonging to the ``metal-rich'' 
component is that of the ``metal-poor'' stars\footnote{Here the Gaussian fit to the metallicity distribution 
at different distances from the centre is only meant to highlight the differences in the 
peak metallicity value between the inner and outer parts.}. We will come 
back to this point in Sect.~\ref{sec:disp_mrmp}. 

Figure~\ref{fig:histo_met} also shows the presence of 31 stars with CaT [Fe/H] $< -2.8$, 
whose presence was not revealed when using the CaT calibration based on 
globular clusters; 11 of these stars have [Fe/H]$<-3$, i.e. they could be classified as extremely metal poor stars. 
The presence of stars of these metallicities excludes the 
hypothesis that Sextans was formed from a pre-enriched medium of [Fe/H]$> -3$ as 
\citet{helmi2006} put forward.

Seven Sextans stars with CaT [Fe/H] $ \la -2.7$ have been followed up 
at high spectral resolution by \citet{aoki2009} and Tafelmayer et al. (2010), 
showing that the CaT calibration holds down to these low metallicities. 

\subsection{Velocity gradients}
Early studies on dSphs, based on datasets of dozens of line-of-sight velocities 
in the central regions of these objects, had not yielded statistically 
significant detections of velocity gradients in these galaxies and therefore 
dSphs have commonly been regarded as exclusively pressure supported systems. 
The gathering of much larger datasets, with hundreds line-of-sight velocities 
spread throughout the face of dSphs, led to 
the detection of velocity gradients in several of these objects.

\begin{figure*}
\begin{center}
\includegraphics[width=0.45\textwidth]{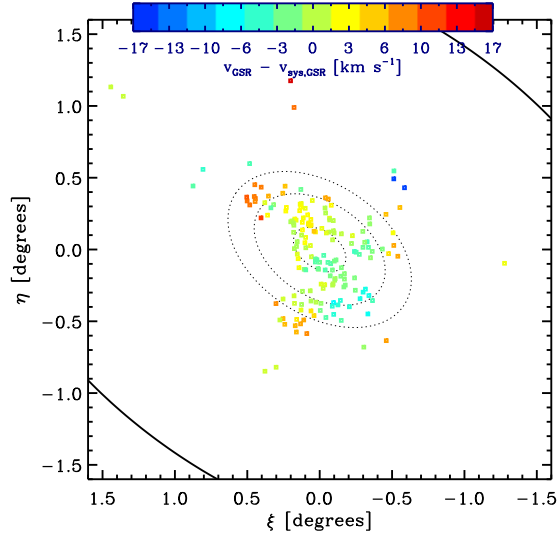}
\caption{Velocity field for probable members to Sextans. The velocities are smoothed with 
a ``median filter'', i.e. we associate to each star 
the median velocity of the stars located within a square of side 0.2 deg centred on the considered star; 
this smoothing is just for visual purposes and aims at reducing the velocity variations due to the dispersion of the galaxy.
We use the velocities in the GSR frame to avoid spurious gradients introduced by the component of the Sun and LSR motions along the l.o.s. to 
Sextans. The colour bar gives the velocity scale. The dotted ellipses are placed at 0.2, 0.5, and 0.7 deg to give an idea of the distance scale. 
The solid ellipse shows the nominal tidal radius \citep[from][]{IH1995}.
}
\label{fig:velocityfield}
\end{center}
\end{figure*}

Figure~\ref{fig:velocityfield} shows a velocity field for Sextans obtained 
from the probable members of the galaxy. There are hints of the presence of a velocity gradient along 
the projected major axis of Sextans, but from this plot it is also evident that, given 
the small number of stars distributed over a large area, more details, such as velocity contours,  
cannot be reliably obtained.

We analyse the velocity trends of the 
probable members along different directions across the face of the galaxy. 
Given the relatively sparse spatial sampling, we explore only 4 directions: along the major and 
minor axes (P.A.$= 56^{\circ}, 146^{\circ}$, respectively) and along 2 intermediate axes 
(P.A.$= 101^{\circ}, 191^{\circ}$). We consider a slit of variable width along those axes and 
explore the trend of the line-of-sight velocity in the GSR frame as a function of the linear distance 
from the centre of the galaxy for those stars within the ``slit''. 
We decided not to bin the data, as they are too sparse.

\begin{figure*}
\begin{center}
\includegraphics[width=0.4\textwidth]{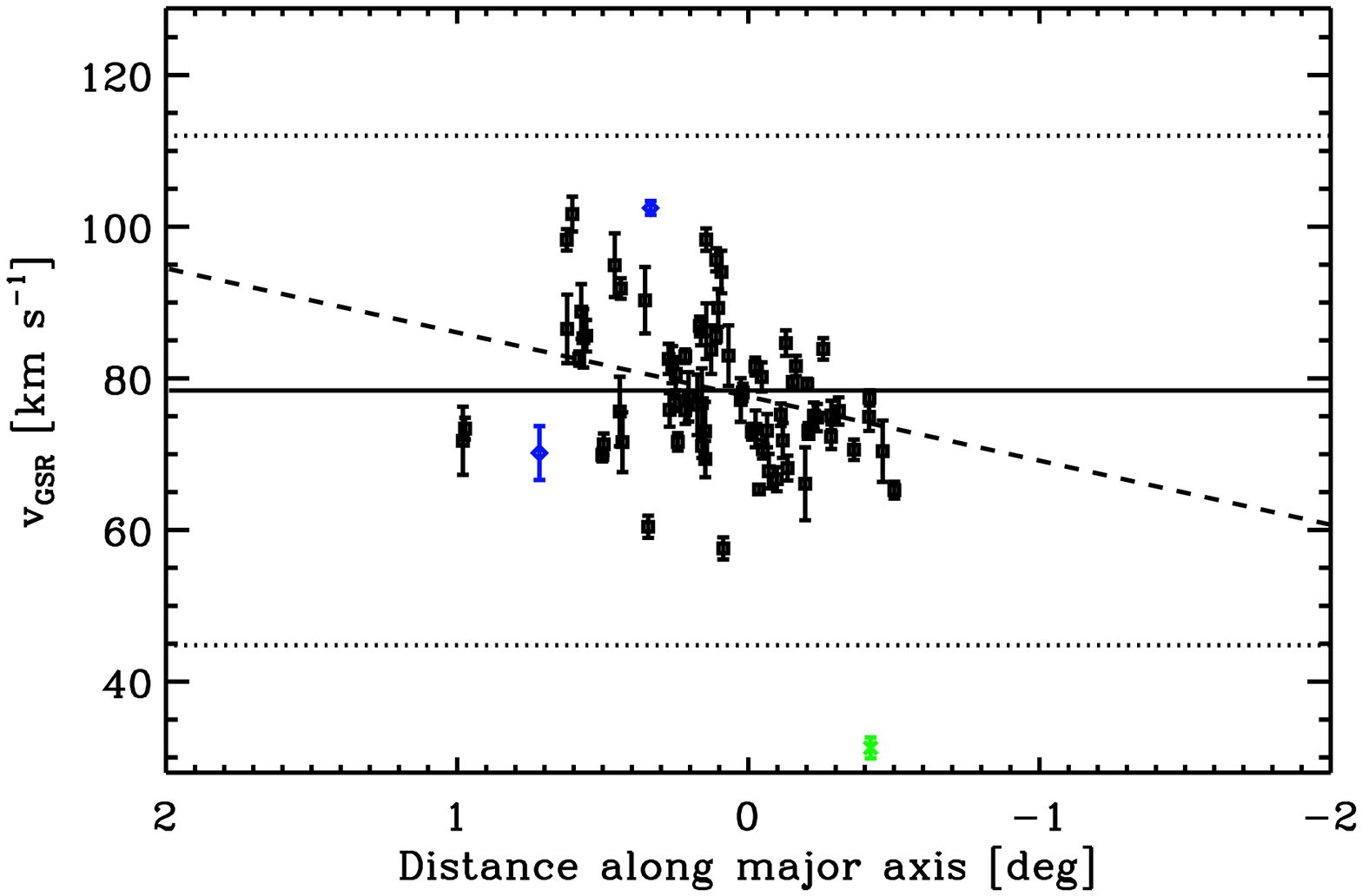}
\includegraphics[width=0.4\textwidth]{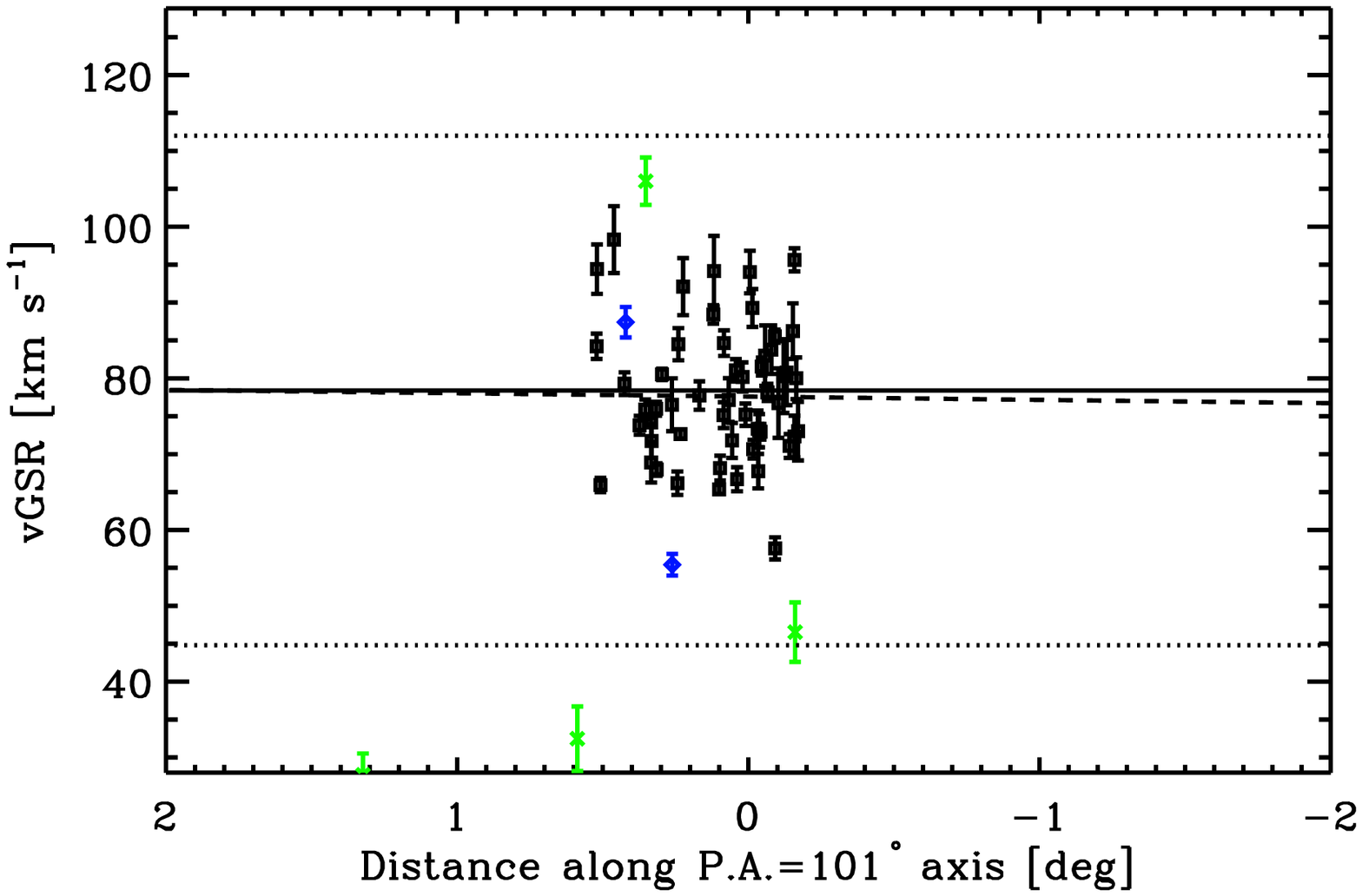}
\includegraphics[width=0.4\textwidth]{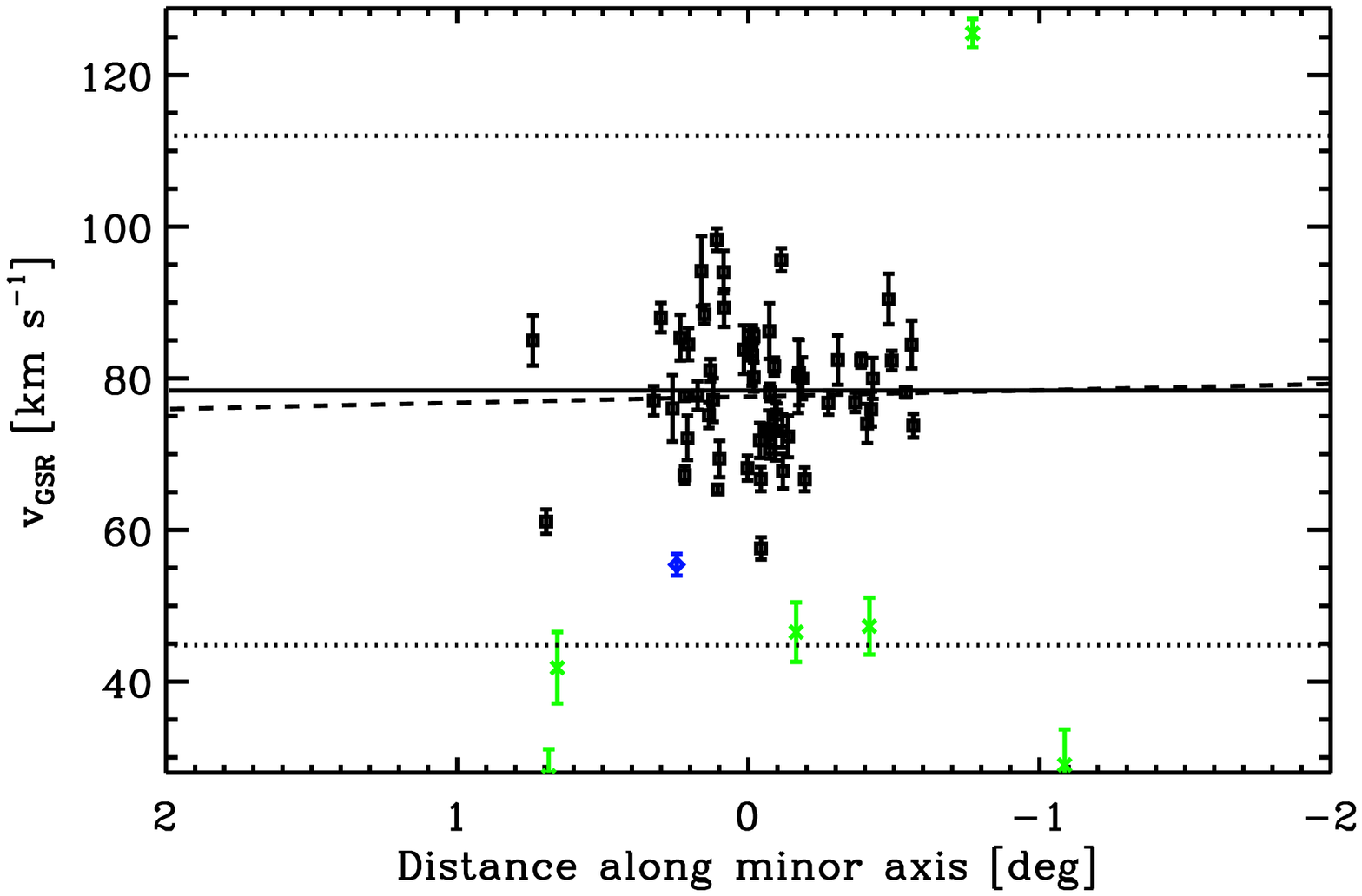}
\includegraphics[width=0.4\textwidth]{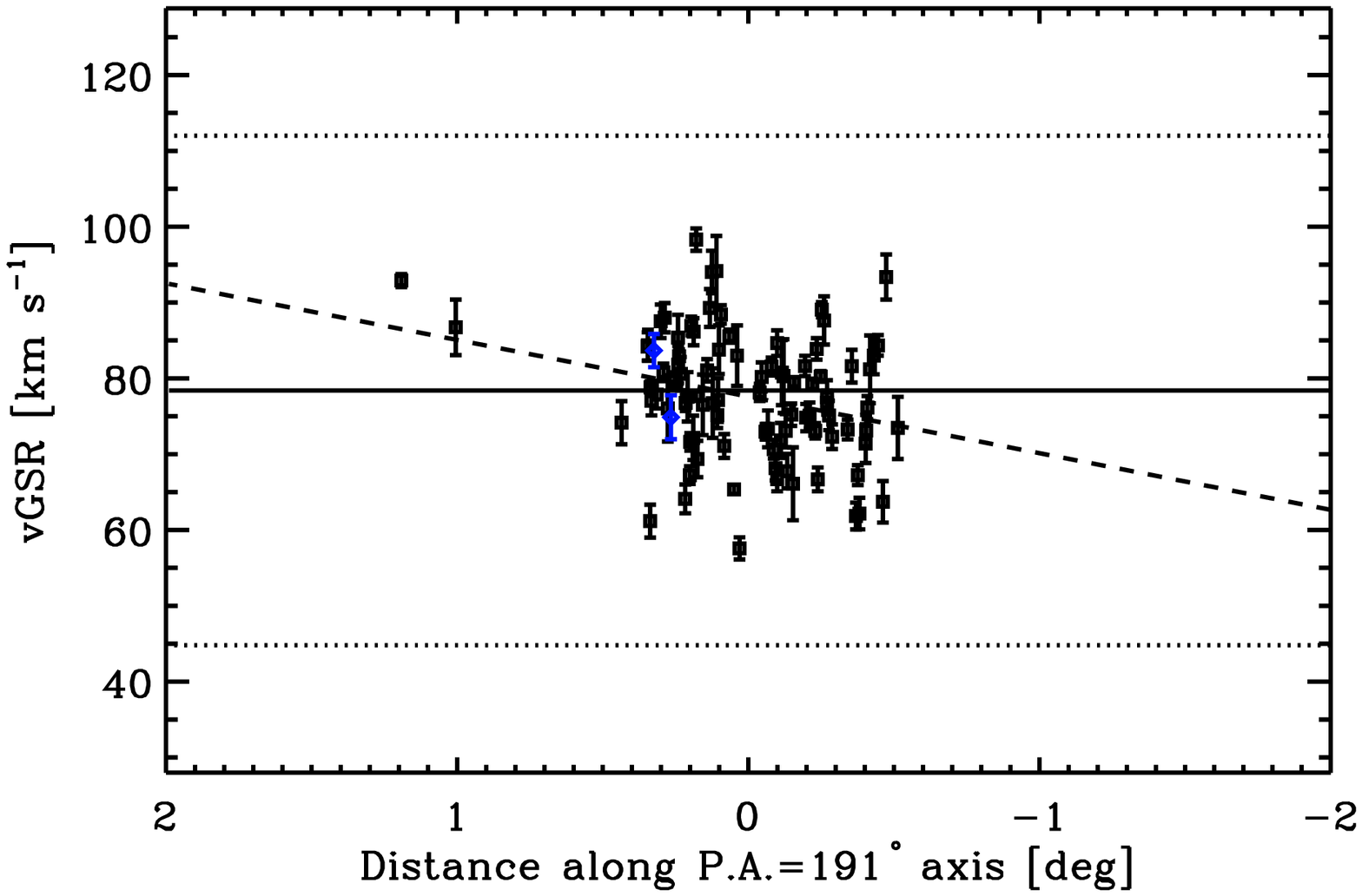}
\caption{$v_{\rm GSR}$ for probable Sextans members versus distance along axes at P.A.$=56^{\circ} \mathrm{(major\ axis)}, 
101^{\circ}, 146^{\circ} \mathrm{(minor\ axis)}, 191^{\circ}$ for stars located within $\pm 0.15^{\circ}$ from these axes 
(squares with error bars). The probable non-members 
within the same spatial region are shown as diamonds and crosses with error-bars (stars with velocity consistent 
with membership but with \mgi EW too large for giant stars and stars with velocity outside the 3$\sigma$ selection 
region, respectively). The solid line indicates the systemic velocity of Sextans; the dotted lines indicate 
a region in velocity of $\pm 5 \sigma$ from the systemic; the dashed line shows the best-fitting straight line 
to the data (see the text for the values of the best fit).}
\label{fig:slit_015}
\end{center}
\end{figure*}

Figure~\ref{fig:slit_015} shows the results for a slit width of 0.3 deg. As fiducial functional form for the gradient 
we use a straight-line, with the intercept fixed to the 
systemic velocity, and we perform an unweighed fit to the data points referring to the probable members. 
In order to derive the error in the fitted slope, we create 10000 mock velocity datasets 
in which the stars are as many as the number of probable members and with the 
same positions, but the velocities are drawn from Gaussians centred on the 
measured velocities and with dispersion given by the velocity errors and the global dispersion of Sextans added in quadrature. For each of these mock 
datasets we repeat the unweighed fit to the velocities along the various axes; 
the error in the slope is given by the 1$\sigma$ level in the distribution 
of fitted slopes. For this slit width the detected 
slopes are: $8.5 \pm 3.0$ \kmsdeg along the major axis,  
$-0.8 \pm 4.2$ \kmsdeg along the minor axis, 
$0.4 \pm 5.3$ \kmsdeg along the axis at P.A.$= 101^{\circ}$ and 
$7.5_{-3.0}^{+3.4}$ \kmsdeg along the axis at P.A.$= 191^{\circ}$. 

In order to assess the significance of such detections we repeat the above procedure 
reshuffling the measured velocities 10000 times, where this should be equivalent to 
eliminating any coherent motion and therefore simulating a non-rotating dSph with the same 
overall dispersion as measured for Sextans. 
We then calculate how many times the mock datasets give amplitudes larger 
than the measured ones. For a slit width of 0.3 deg the percentages are  
0.4\%, 58.4\%, 46.9\% and 0.9\% along the major, minor, P.A.$= 101^{\circ}$ and P.A.$= 191^{\circ}$ axes, 
respectively.\footnote{We checked whether the gradient along the P.A.$= 191^{\circ}$ axis 
may be driven by the two points at distance larger than 0.9 deg, by repeating the determination removing 
these two points. The gradient is still present, $6.3 \pm 3.7$ \kmsdeg, although the significance 
slightly decreases (the percentage of mock dataset giving as large a gradient increases to 4.5\%).}. 
The only statistically significant detections appear to be the ones  
along the major axis and along the axis at P.A.$= 191^{\circ}$. 
If these velocity gradients 
are due to rotation about the minor axis of Sextans, we would expect to detect similar velocity gradients 
along the P.A.$= 101^{\circ}$ and P.A.$= 191^{\circ}$ axes, given that they are symmetric about the minor axis; 
however, while the spatial coverage along the P.A.$= 191^{\circ}$ is quite homogeneous and symmetric 
about the centre of the galaxy, this is not the case along the P.A.$= 101^{\circ}$ axis, along which 
we have data only in the N-W part of the galaxy; it is therefore unclear if the discrepancy in 
the detection is due to the difference in spatial coverage.  

We find that, when increasing 
the slit width, for example of 0.4 deg, the amplitude of the gradients slightly decreases, 
which suggests that the kinematics along a certain axis starts to be contaminated by the inclusion of more stars 
moving differently because placed at different distances from the centre. In this case the percentage of the 
mock datasets giving amplitudes larger than the measured ones is larger than for a slit width of 0.3 deg, 
but the gradients along the projected major axis and the P.A.$= 191^{\circ}$ axis remain statistically significant.

Another concern in the derivation of the velocity gradient is the possible inclusion in the 
sample of 
contaminants from the Milky Way, or vice versa the possible exclusion of Sextans stars 
based on the 3$\sigma$ cut-off. However, none of the two points appear to be an issue in 
the analysis here carried out as it is visible from Fig.~\ref{fig:slit_015} where we 
also show the non-members falling in the considered slits: from these figures 
it is clear that the fraction of unidentified contaminants still present in the 3$\sigma$ velocity range 
is likely to be negligible, and also that we are unlikely to have missed probable 
members because of the 3$\sigma$ cut-off.

As mentioned already, with the current spatial sampling it is difficult to extract a velocity field and 
to determine a rotation pattern. We adopt a simple empirical approach and assume that $v_{\rm rot} = k * d_{\rm min}$, 
where $k = 8.5$ \kmsdeg and $d_{\rm min}$ is the angular distance of each star from the minor axis along a direction 
parallel to the major axis (with $d_{\rm min}$ positive above the minor axis and negative below the minor axis, 
the formula gives receding and approaching velocities above and below the minor axis, respectively). 
To assess how good a representation of the data this is, we subtract the $v_{\rm rot}$ from the stars 
classified as probable members, and re-derive the velocity field and the trends along the various slits.
There appears to be a residual pattern in the velocity field from the ``rotation-subtracted'' velocities; 
indeed, velocity gradients are still present along the various axes, but they do not appear to be statistically significant. For a slit width of 0.3 deg the detected 
slopes are: $-0.1 \pm 3.1$ \kmsdeg along the major axis,  
$-1.8_{-3.9}^{+4.5}$ \kmsdeg along the minor axis, 
$7.1_{-5.5}^{+5.2}$ \kmsdeg along the axis at P.A.$= 101^{\circ}$ and 
$1.6 \pm 3.1$ \kmsdeg along the axis at P.A.$= 191^{\circ}$. 
The percentages of mock datasets that give amplitudes larger 
than the measured ones are large, 
48.4\%, 66.4\%, 9.0\% and 30.6\% along the major, minor, P.A.$= 101^{\circ}$ and P.A.$= 191^{\circ}$ axes, respectively, 
i.e. the gradients do not appear to be highly statistically significant, meaning that the rotation 
pattern we assume is a reasonable representation of the data.  

It appears that a weak velocity gradient is present in Sextans, but the direction and 
amplitude of this gradient is difficult to constrain. Note that our determination differs from 
the determination by \citet{walker2008}, who find a gradient of $-2.1 \pm 0.8$ \kmsdeg along 
a P.A.$= 120^{\circ}$. Given that the velocities between the two studies compare 
well (see Appendix), this difference is most likely due to the different coverage 
and size of the sample, as well as the methodology used. A 
larger statistical sample and coverage of the galaxy would help quantifying this better; 
however, given the low surface brightness and large extent on 
the sky of Sextans, this may be a challenging task. 

\subsection{Observed velocity dispersion profile}

\subsubsection{All members} \label{sec:disp_all}
We derive the l.o.s. velocity dispersion profiles for Sextans by binning the velocities of stars 
at similar radii using the sample of probable 
members as determined in Sect.~\ref{sec:finalmem} and deriving the average velocity and dispersion per bin 
using the procedure described in Hargreaves et al. (1994). 

In deriving the l.o.s. velocity dispersion profile we checked for the effect of a number of issues.

First, we check the effect that the number of stars per bin has on the derived 
l.o.s. velocity dispersion profile by fixing the 
number of stars per bin to a constant value of N$=$20,25,30 and 35 (except for the last bin, 
which is allowed to have a lower number of points). We find that, even though the 
shape and amplitude of the l.o.s. velocity dispersion profiles are consistent 
within the error-bars among all the cases, the binning with N$=$35 stars is the most reliable one 
because it produces 
less oscillations in the trend of 
average velocities per bin, and that these velocities are consistent with the systemic velocity. 

\begin{figure*}
\begin{center}
\includegraphics[width=0.45\textwidth]{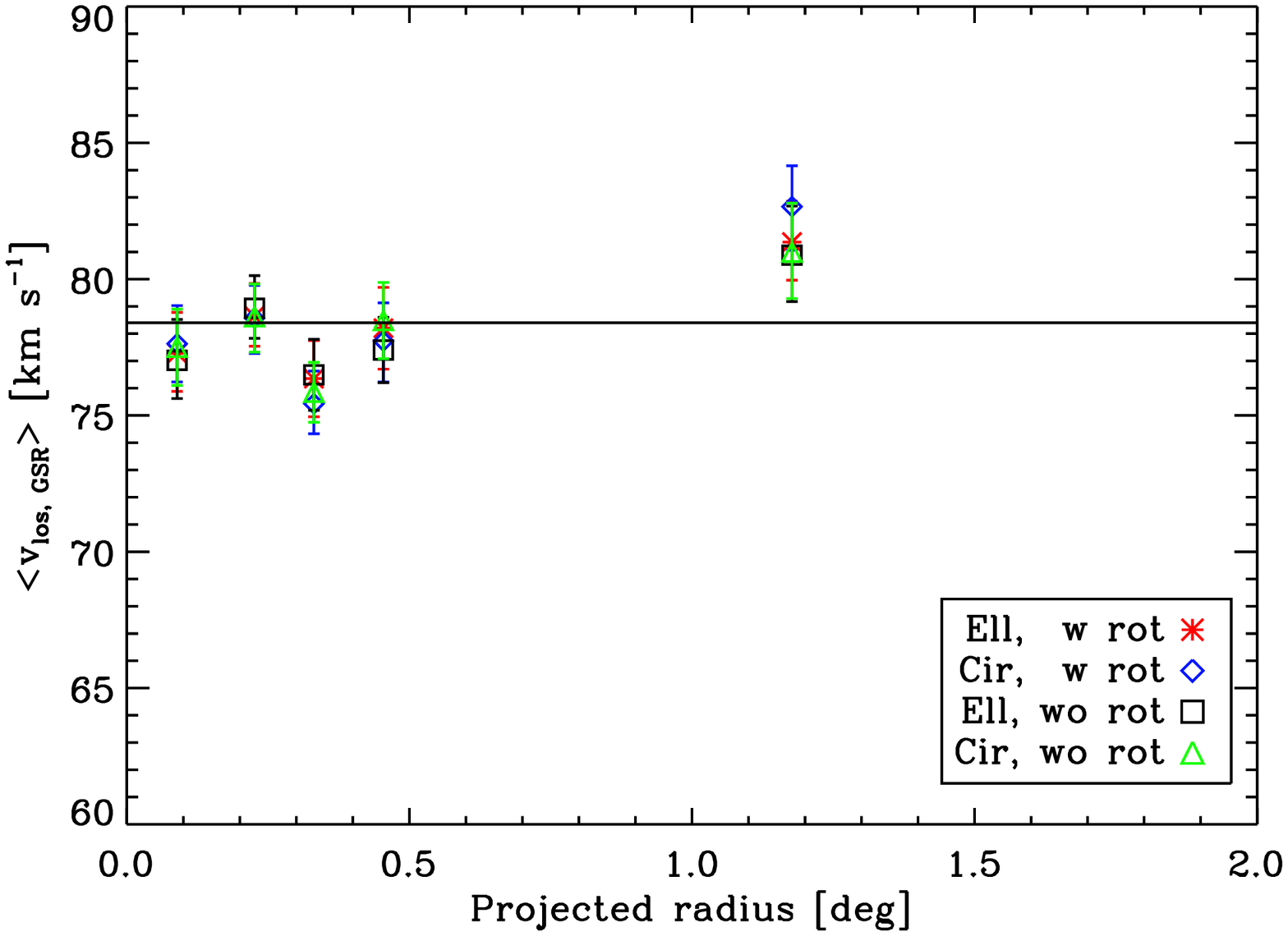}
\includegraphics[width=0.45\textwidth]{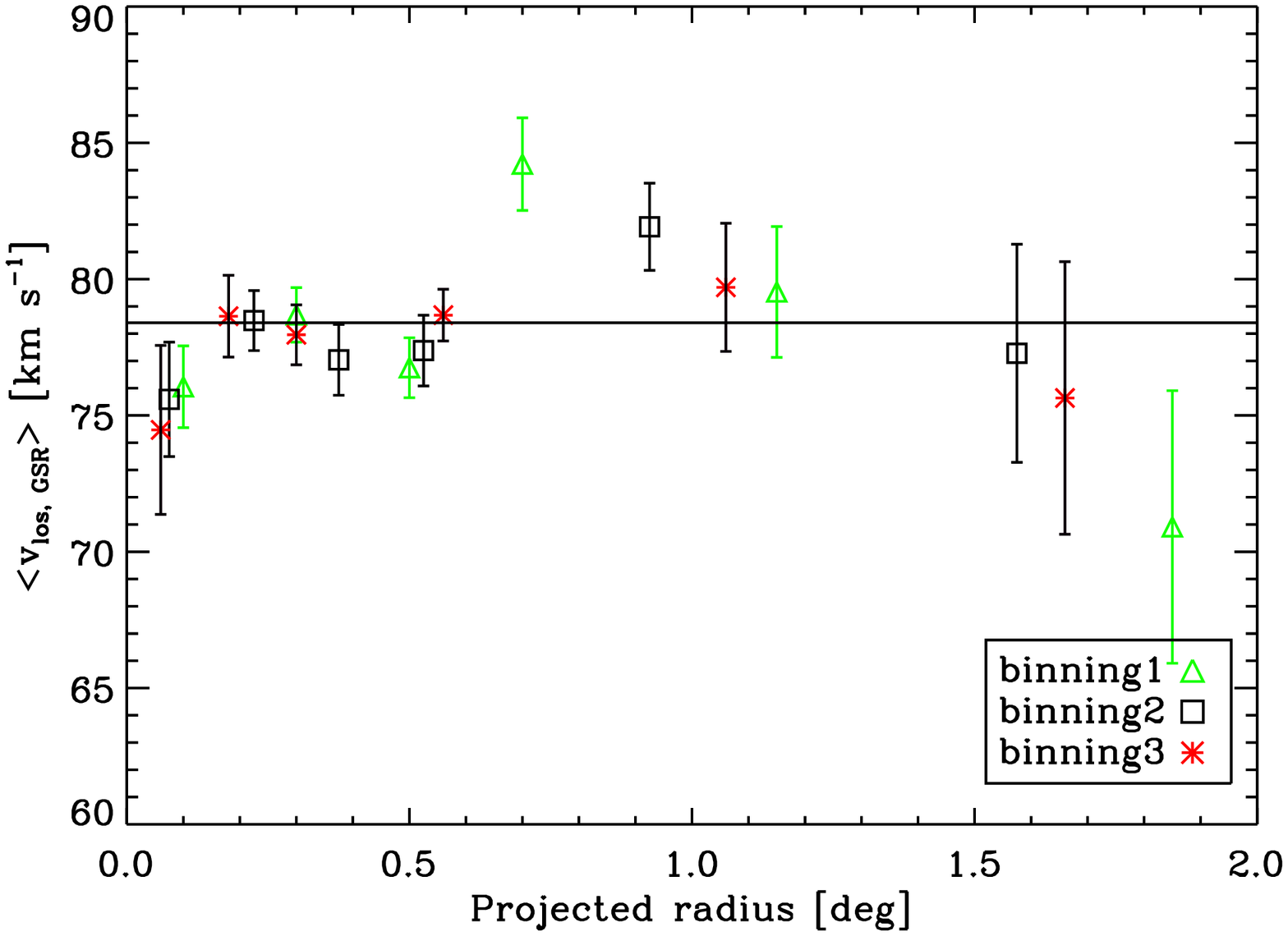}
\includegraphics[width=0.45\textwidth]{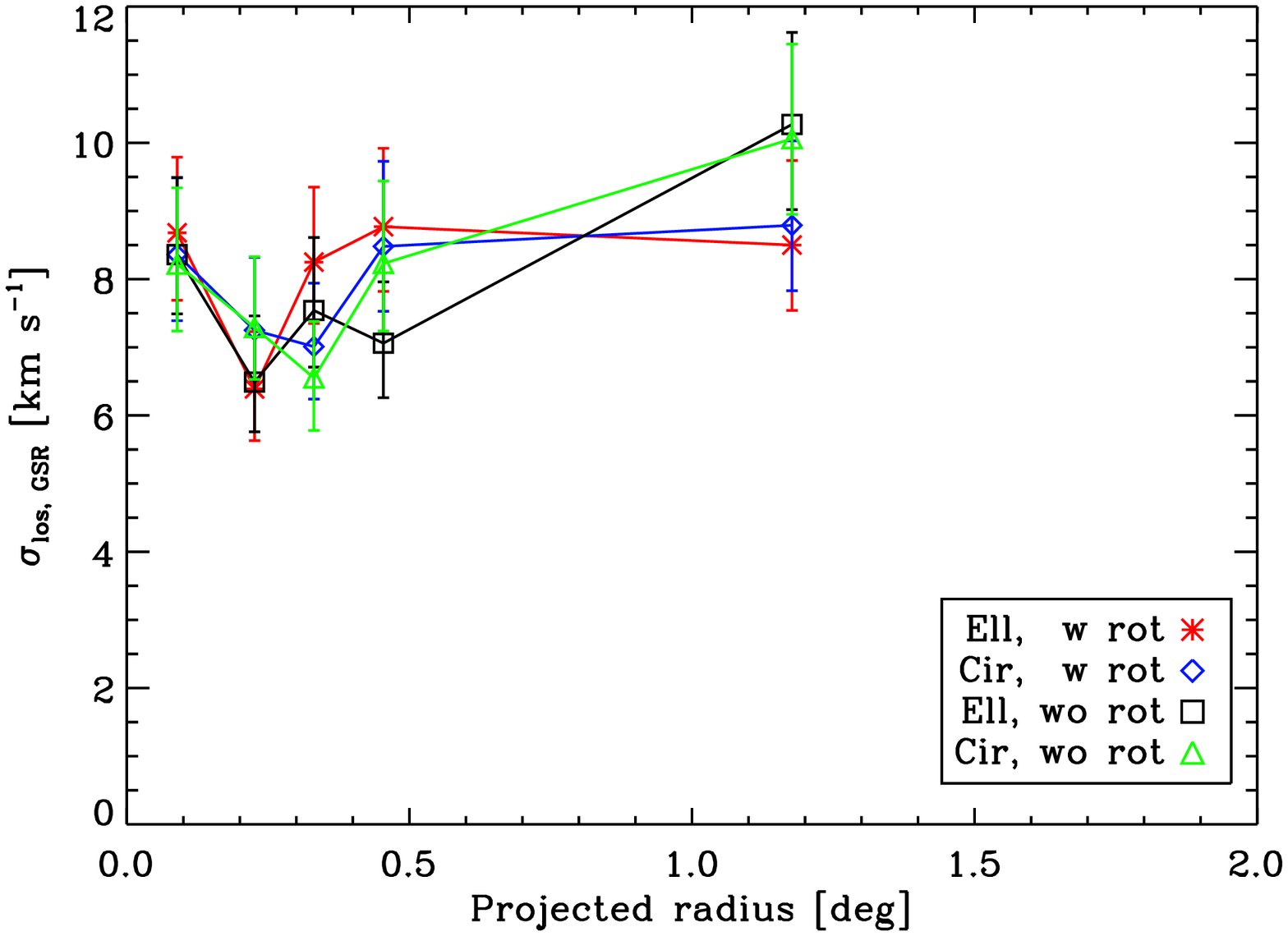}
\includegraphics[width=0.45\textwidth]{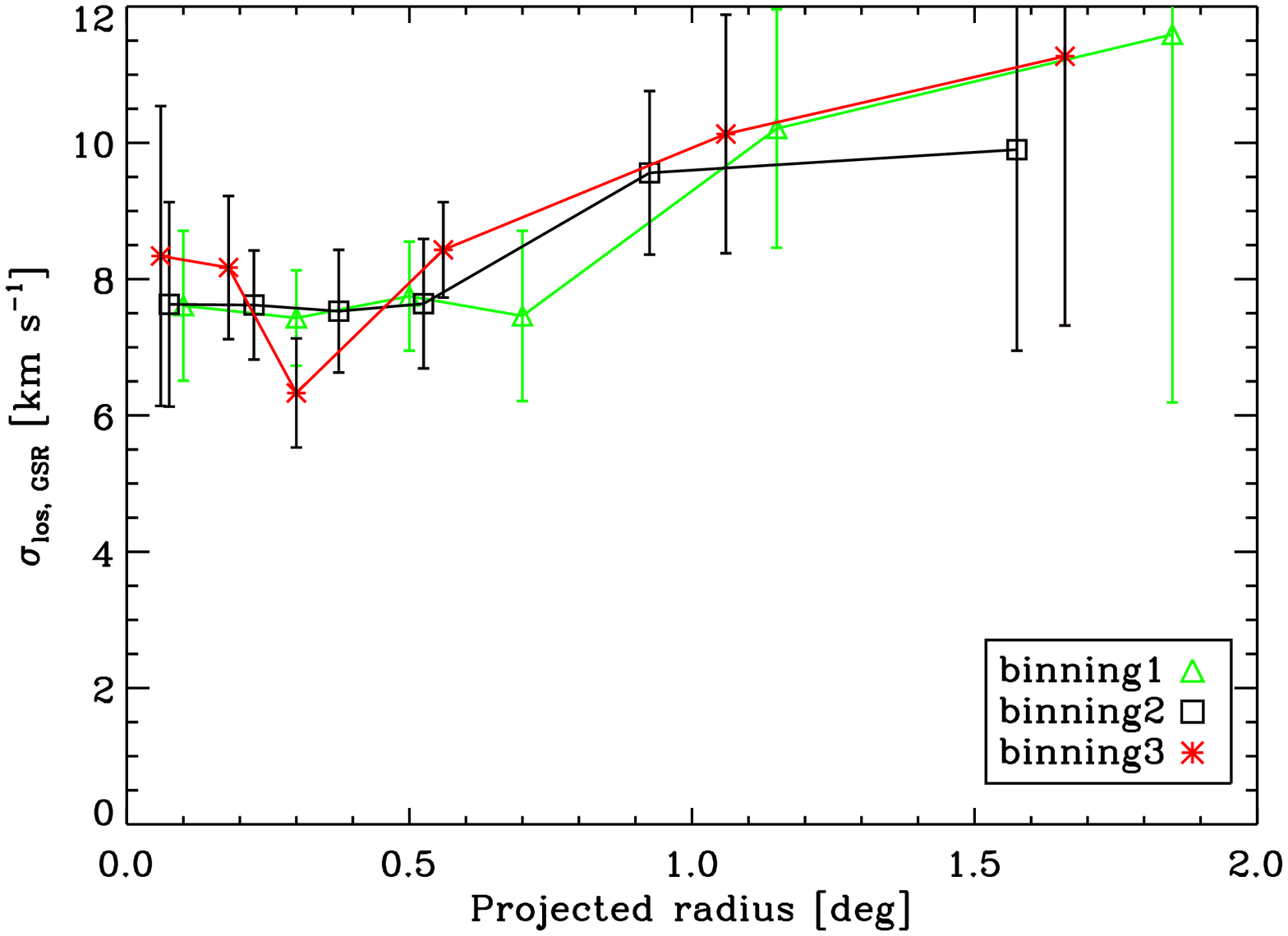}
\caption{L.o.s. average velocity (top) and velocity dispersion profile (bottom) 
in GSR system for probable Sextans members versus projected radius. 
Left: profiles derived keeping the number of stars per bin 
approximately constant to 35 stars per bin; the profiles have been derived both when the velocity gradient is included (asterisks: elliptical binning; 
diamonds: circular binning) and when it is subtracted from the individual velocities 
(squares: elliptical binning; triangles: circular binning). 
Right: profiles derived for 3 different choices of binning with bins of increasing width with projected radius, 
in the case  
when the gradient is subtracted from the individual velocities and elliptical binning. 
The horizontal solid line in the top panel 
indicates the systemic velocity of Sextans.
}
\label{fig:disp_3sigma_all_harg_GSR}
\end{center}
\end{figure*}

We then explore the effect of binning the data in elliptical 
annuli (of constant ellipticity and position angle 
($e=0.35$ and P.A.$= 56^{\circ}$) and circular annuli, using a constant number of N$=35$ stars per bin. 
In fact, one uses stars 
with a non-spherical spatial distribution as kinematic tracers of an underlying distribution 
which may be spherical (as well as oblate, prolate or even triaxial). We also use 
a constant number of N$=35$ stars per bin 
when exploring the effect of subtracting the detected velocity gradient onto the 
determination of the velocity dispersion profile (hereafter we refer to 
the profile derived from the individual velocities from which the 
velocity gradient has been subtracted as the ``rotation subtracted'' profile). These 
results are shown in Fig.~\ref{fig:disp_3sigma_all_harg_GSR} (left): the l.o.s. 
velocity dispersion profiles derived from the elliptical and circular binning, 
when subtracting or not rotation, all agree with each others within the error-bars. 
In the following we adopt the rotation subtracted velocities and the 
elliptical binning. 

The effect of using bins of variable width which increases with 
projected radius, instead than using a constant number of stars per bin, 
is shown in the right-hand panel of Fig.~\ref{fig:disp_3sigma_all_harg_GSR}. 
We explored 3 different binnings, which are labelled as ``Binning1'', ``Binning2'' 
and ``Binning3'' in the right-hand side panel of Fig.~\ref{fig:disp_3sigma_all_harg_GSR}). 
All these choices of binning are compatible 
with the trend shown when keeping the number of probable members per bin constant. 
In the following, 
When deriving the mass profile of Sextans we will use the l.o.s. velocity dispersion profile 
from ``Binning2''; however, the effects of a different binning on the mass determination will also 
be considered, by checking the results obtained when using the profile from ``Binning3''. 
We will not use ``Binning1'' for the analysis because it represents 
the case where the average velocity per bin differs the most from the systemic. 

The l.o.s. velocity dispersion profile of Sextans appears to 
slightly increase with radius although, given the large size of the error-bars 
in the last bins, it is also compatible with remaining constant. 

To compare the l.o.s. velocity dispersion profile  published by \citet{walker2007} 
and ours we should consider the case of the circular binning, with no velocity gradient 
subtracted: in this case our profile remains approximately constant around 8 \kms, consistent 
with the determination by \citet{walker2007}.

\subsubsection{``Metal-rich'' and ``Metal-poor'' stars} \label{sec:disp_mrmp}
In previous work, we have combined the information on the 
spatial location, metallicity and velocity of individual stars in dSphs. This allowed us to explore 
links between the metallicity and the kinematics of stars 
in the Sculptor and Fornax dSphs (T04 and B06, respectively), and shown that in these systems 
metal-rich stars - centrally concentrated - have a smaller velocity dispersion 
than metal-poor stars - which in turn have a more extended spatial distribution. The metallicity 
chosen to illustrate this different kinematic behaviour between the colder and hotter stellar components  
is somewhat subjective, but it is chosen on the basis of considerations on the metallicity 
distribution in order to minimize contamination 
between the two metallicity components. 

As seen in Sect.~\ref{sec:met} at R$<$0.8 deg, the region where the whole range 
of metallicities is present, the metallicity distribution of Sextans is well-approximated with 
the sum of two Gaussians, a ``metal-rich'' and a ``metal-poor'' one, which cross 
at [Fe/H]$\sim -2.3$. Therefore to minimize contamination between these two components, 
we exclude the range $-2.4 < $ [Fe/H] $< -2.2$ and we derive 
the trend of mean velocity and dispersion for the stars more metal-rich than [Fe/H] $= -2.2$ and 
more metal-poor than [Fe/H] $= -2.4$ over different spatial bins. 

Figure~\ref{fig:chemodyn} shows that also in Sextans ``metal-rich'' and ``metal-poor'' stars 
display a different behaviour (akin to our findings in the Fornax and Sculptor dSphs), with the former population 
in general kinematically colder than the latter. The profile of the metal-rich stars appears 
to be slightly declining, while for the metal-poor stars it is everywhere constant 
(except in the first bin, see next section) and extends further than the profile for the MR stars, as 
a consequence of the larger spatial extent of the metal-poor stars. It remains to be proven if the different spatial extent of the 
stellar population analysed is the only factor that causes the different kinematics, 
or if one needs to take into account also different orbital properties for the 
different stellar populations. Also, the outer bin for the ``metal-rich'' stars 
suffers from low number statistics and would certainly benefit from a larger sample of 
stars with [Fe/H]$> -2.2$.

\subsection{Substructures}
\citet{kleyna2004} detected a kinematically cold substructure, with dispersion close to zero, from 7 stars with radial 
velocity measurements in the inner 5 arcmin of Sextans. One of the hypothesis that they put forward is that 
the stars giving raise to this kinematic feature may belong to the remnant of a disrupted stellar 
cluster which spiraled into Sextans centre. The authors note that this hypothesis can account for the 
sharp central rise in the light distribution of Sextans (e.g. Irwin \& Hatzidimitriou 1995). Also 
the distribution of blue stragglers in Sextans, with the 
brighter (i.e. more massive) blue stragglers more centrally concentrated than the fainter ones \citep{lee2003} could 
be explained by this mechanism. The authors estimate a luminosity of 1.3$\times 10^5$ \sL for the 
disrupted cluster. The existence of this central cold substructure 
has not been confirmed by subsequent studies such as \citet{walker2006}, who instead 
find another cold substructure of estimated luminosity $\sim 10^4$ \sL\, at a different location, i.e. 
around the core radius of the galaxy.

\begin{figure*}
\begin{center}
\includegraphics[width=\textwidth]{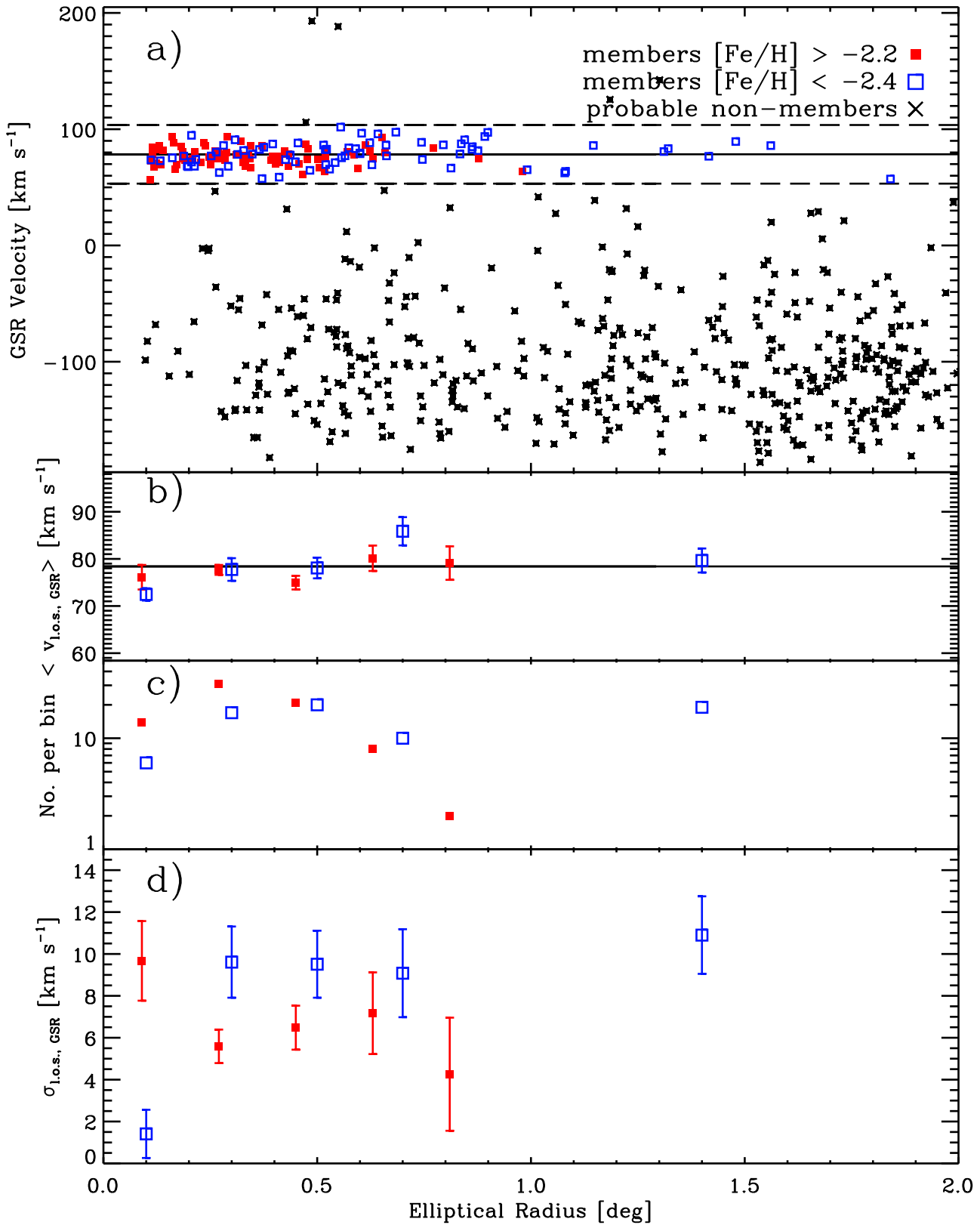}
\caption{Kinematic properties for probable Sextans members more metal rich than [Fe/H]$=-2.2$ 
(red filled squares) and more metal poor than [Fe/H]$=-2.4$ (blue open squares). 
We show the variation of the rotation-subtracted GSR velocity versus elliptical radius (a); 
average rotation-subtracted GSR velocity (b); number of stars (c); velocity dispersion 
profile using rotation-subtracted GSR velocities (d). The solid horizontal line indicates the systemic 
velocity, the horizontal long-dashed lines show the region used for the 3$\sigma$ membership selection. 
}
\label{fig:chemodyn}
\end{center}
\end{figure*}

The inner point in Fig.~\ref{fig:chemodyn} shows peculiar characteristics for the metal-poor stars: 
the velocity dispersion is very cold, 1.4$\pm$1.2 \kms, and the average velocity is 
72.5$\pm$1.3 \kms, about 4$\sigma$ away from the systemic\footnote{Here with $\sigma$ 
we denote the error in the determined average velocity for the substructure.}. In the binning used 
these values are derived from the 6 innermost metal-poor stars, however from panel (a) of Fig.~\ref{fig:chemodyn} 
it can be seen that other 3 MP stars at R$<$ 0.22 deg share very similar velocities to 
this group of stars; only 1 MP star at R$<$ 0.22 deg is found at very different velocities and 
can therefore be thought of as belonging to the main Sextans population. This group of 9 stars do not only share very similar 
distances and kinematics but also metallicities, with an average metallicity of 
[Fe/H]$= -2.6$ and a scatter of 0.15 dex. Note that the average 
metallicity error for the stars in the substructure is 0.29 dex, much larger than 
the measured scatter in the metallicity, which means that the intrinsic scatter is 
going to be much lower than 0.15 dex. The similar metallicity of the stars found in the 
cold kinematic substructure would point to these stars previously belonging to a single stellar population,
i.e. a stellar cluster. 

Note that the information on metallicity is useful for such kind of detections: 
when considering all the stars together in the inner bin, with no distinction 
made on metallicity, the velocity dispersion is much larger, consistent with the 
rest of the galaxy. 

Assuming that the ratio of stars in the substructure (9) with respect to the 
total number of probable Sextans members (174) is representative, 
then the substructure would account for 5\% of the overall Sextans population. 
Assuming a luminosity for Sextans from Table~\ref{tab:par} and a 
stellar mass-to-light ratio (M/L)$_{\rm lum,V} = 2$ typical of globular clusters 
\citep{illingworth1976, pryor1988} as plausible for an old stellar population, 
then crude estimates of the total luminosity and mass of this structure 
are $2.2 \times 10^4$ \sL and $4.4 \times 10^4$ \sm. If this substructure is the same 
one found in \citet{kleyna2004}, our estimated luminosity is approximately one 
order of magnitude smaller than what \citet{kleyna2004} estimated, that is $1.3 \times 10^5$ \sL.  
However the estimate was carried out in a different 
way, where  \citet{kleyna2004} assumed that all the light in the central 10 arcmin 
is due to cluster debris. 

We can instead assume that the estimated light in the central 
10 arcmin, $1.3 \times 10^5$ \sL, is actually given by the sum of the light from the substructure plus  
the light from the overall Sextans population. We then use the value of the central surface brightness 
from \citet{IH1995} - derived excluding the central point giving raise 
to the sharp central rise in the light distribution of Sextans - to estimate the contribution of the 
overall Sextans population to the light in the central 10 arcmin: this is about $7.8 \times 10^4$ \sL. 
Therefore the contribution of the substructure would then amount to $5.2 \times 10^4$ \sL, much closer to our 
estimate. It is therefore possible that the substructure here detected is the same one as in 
 the work of \citet{kleyna2004}.

The properties we find are compatible with other cold substructures found 
in other dSphs, such as in Ursa Minor by \citet{kleyna2003}, in Sculptor 
\citep{battaglia2007} and at projected distances of about 1 core radius in Sextans \citep{walker2006}. 
This luminosity corresponds to an absolute M$_V = -6.2$, which is similar to globular clusters 
found in the Fnx, Sagittarius dSphs and in other dwarf galaxies \citep[e.g., see][]{vandenbergh2006}.

To date, the most metal-poor known globular cluster resides in the Fnx dSph, 
with high resolution measurements yielding a [Fe/H] $= -2.5$ \citep{letarte2006}. 
If the stars in the cold substructure here detected do indeed belong to the remnant 
of a disrupted globular cluster, the [Fe/H] values we measure would place it 
among the most metal poor globular clusters known. Given the large 
error-bars of the individual CaT measurements for these stars, 
determinations from high resolution spectroscopic measurements would be needed 
to place this on a secure foot. Detection of the O-Na anti-correlation would 
confirm the hypothesis.

Note that the CaT [Fe/H] 
values are calculated assuming that the stars in the substructure are 
at the same distance of Sextans. If those stars are actually an external 
feature, then the [Fe/H] values would not be correct. However, given the 
average velocity of the substructure well within 1$\sigma_{\rm GSR}$ from the 
systemic velocity of Sextans, it is likely that these stars do indeed belong to the dSph.

\section{Mass determination} \label{sec:mass}
We determine the mass of Sextans by means of Jeans modeling of a spherical and stationary system 
assuming different dark matter density profiles.

The methodology we use consists of comparing the observed l.o.s. velocity dispersion 
$\sigma_{\rm los}$ for each distance bin
with that predicted from the Jeans modeling for different models of the dark matter density profile 
and hypothesis on the velocity anisotropy of the tracer\footnote{By ``tracer population'' 
we mean those objects whose kinematics can be used to recover properties of the total potential. In this case 
our tracers are the spectroscopically observed red giant branch stars.} (see Sect.~\ref{sect:predicted_sigmalos}). 
We explore the space of parameters 
which define each model and determine the $\chi^2$ as:
\begin{equation} \label{eq:chisquare}
\chi^2 = \sum_{i=1}^{Nbins} \biggl(\frac{\sigma_{{\rm los}_i} - 
\sigma_{\rm los}(R_i;
p_{\beta}, p)}{\epsilon_i}\biggr)^2.
\end{equation}
The variable $p$ denotes a characteristic mass parameter of each
dark matter model, $p_{\beta}$ denotes a parameter describing the behaviour of the 
velocity anisotropy (see Sect.~\ref{sec:res_mass}). Finally, $\epsilon_i$ is the
error in the observed l.o.s. velocity dispersion. The best-fitting 
parameters are defined as those for which $\chi^2$ is minimized. 
We quote as errors in the individual parameters 
the projections of the $\Delta \chi^2=2.3 $ region (corresponding to the region of 68.3\% {\it joint} probability 
for a two free parameters $\chi^2$ distribution).

\subsection{Predicted velocity dispersion profile} \label{sec:predicted}
\label{sect:predicted_sigmalos}
The Jeans equation for a stationary spherical system in absence of net streaming motions in any of the directions is \citep{binney1987}:
\begin{equation}
\frac{1}{\rho_*}\frac{d({\rho_* \sigma_{r,*}^2})}{dr} + \frac{2\beta
\sigma_{r,*}^2}{r} = -\frac{d\phi}{dr} = - \frac{V_{\rm c}^2}{r}%
\label{eq:jeans}
\end{equation}
where $\rho_*$ is the density of the tracer; $\beta$ is the velocity anisotropy parameter, defined
as $\beta = 1 -\  \sigma_{\theta,*}^2 / \sigma_{r,*}^2$, assuming
$\sigma_{\theta,*}^2=\sigma_{\phi,*}^2$; $\sigma_{r,*}$, $\sigma_{\theta,*}$ and $\sigma_{\phi,*}$ are the 
velocity dispersions in the ($r, \theta, \varphi$) direction respectively; $\phi$ and $V_{\rm c}$ are the potential and the circular velocity 
of the total mass distribution. Note that $\beta = 0$ if the
velocity ellipsoid is isotropic, $\beta = 1$ if the
ellipsoid is completely aligned with the radial direction, and 
$\beta < 0$ for tangentially anisotropic ellipsoids.

The quantity to compare to the observations is the l.o.s. velocity dispersion of the tracer population \citep{binney1982}:
\begin{equation}
\label{eq:jeans_binneymamon1982}
\sigma_{\rm los}^2(R)= \frac{2}{\Sigma_*(R)} \int_R^{\infty} \frac{\rho_*(r) \sigma_{r,*}^2 \ r}{\sqrt{r^2-R^2}} 
(1 - \beta \frac{R^2}{r^2} ) dr 
\end{equation}
where $R$ is the projected radius (on the sky) and $\Sigma_*(R)$ is the mass surface density 
of the tracer.

We refer to \citet{mamon2005} for the derivation of the l.o.s. velocity 
dispersion profile from the Jeans equation using different hypotheses on $\beta$.

As the discovery of multiple 
stellar populations in dSphs is a recent one, traditionally dSph galaxies 
have been treated as single component systems. In \citet{battaglia2008b} we 
modelled each stellar component separately in the Sculptor dSph, showing that this 
allows to relieve some of the degeneracies. This kind of analysis needs imaging such that 
the surface number density profiles of the different components can be 
accurately determined. This is not the case to date for Sextans, therefore 
until better photometric data will come we will 
treat all the stars in Sextans as in one component. 

Below we discuss in detail 
the ingredients to solve the Jeans equation.

$\bullet$ {\bf The spatial distribution of the tracer:} 
\citet{IH1995} found that the surface brightness profile 
of Sextans is best fit by an exponential profile 
with scale radius $R_e=15.5'$. This is equivalent to assuming a 
Sersic profile with Sersic radius $R_S = 15.5'$ and shape parameters $m =1$ as done 
by \citet{walker2007} and \citet{lokas2009}.  
For the mass-to-light ratio we 
use (M/L)$_V = 1$ (M/L)$_{V,\odot}$ and luminosity and distance from Table~\ref{tab:par}.

The 3D density profile is derived from the surface brightness profile through 
inversion of Abel integrals, assuming that the stars are spherically distributed. 
We refer to \citet{lokas2005} for all the relevant formulas.

$\bullet$ {\bf The kinematics of the tracer - Velocity gradient:}
 The expression for the projected velocity dispersion (Eq.~\ref{eq:jeans_binneymamon1982}), 
which measures the projected random motion in a galaxy, has been derived 
for the hypothesis that the system is not rotating. 
In systems where the rotation law can be accurately derived and there is enough 
statistics to derive the dispersion profile 
along both the major and minor axes, one can use the Jeans 
equations in cylindrical coordinates. The present coverage 
of this dataset does not allow this alternative. Therefore we use the 
observed l.o.s. velocity dispersion profile derived from the rotation 
subtracted individual velocities. 

$\bullet$ {\bf The kinematics of the tracer - Velocity anisotropy:} 
Since the variation of the velocity anisotropy with radius is not known 
(as this requires proper motions), 
we consider two hypotheses: $\beta$ constant with radius; and using the Osipkov-Merritt parametrization for $\beta$ 
\citep{osipkov1979, merritt1985} 
\begin{equation}
\label{eq:beta_om}
\beta_{\rm OM} = r^2/(r^2 + r_a^2)
\end{equation}
where $r_a$ is the anisotropy radius. 
In the Osipkov-Merritt parametrization the anisotropy is always $\ge 0$, i.e. it is never tangential. 
The central regions are isotropic, and for $r_a \to \infty$ the anisotropy 
becomes purely radial. At $r= r_a$, $\beta=0.5$. 
The smaller $r_a$, the faster the anisotropy becomes very radial. Models with large $r_a$ correspond 
to models with almost isotropic behaviour.

$\bullet$ {\bf Total mass distribution:}
Since from previous works the contribution of the stars to the overall potential has been shown to be 
negligible in Sextans and in dSphs in general, we consider the dark matter 
halo as the only contributor to the kinematics of the tracer population.

We consider two different models for a spherically symmetric 
dark-matter halo potential:

- {\it Pseudo-Isothermal sphere.} This model has been extensively
used in the context of extragalactic rotation curve work 
\citep[see][and references therein]{swaters2000}.  The density
profile associated to this model is:
\begin{equation}
 \rho(r) = \rho_0\frac{r_{\rm c}^2}{(r_{\rm c}^2 + r^2)},
\end{equation}
where $r_{\rm c}$ is the core radius, $\displaystyle \rho_0 = \frac{V_{\rm
c}^2(\infty)}{4 \pi G r_{\rm c}^2}$ is the central density and $V_{\rm c}(\infty)$ 
is the asymptotic circular velocity. At large radii the
density behaves as $\rho\propto r^{-2}$.

The resulting mass profile is:
\begin{equation}
\label{eq:iso_massprofile}
M(<r) = 4 \pi \rho_0 r_{\rm c}^2 \bigl(r - r_{\rm c} \arctan \frac{r}{r_{\rm c}}\bigr).
\end{equation}
The profile is completely defined by $\rho_0$ and $r_c$, or any 
couple of non-degenerate parameters.

- {\it NFW model.} This profile is motivated by cosmological 
N-body simulations in a CDM framework \citep{nfw1996, nfw1997}. 
In this case the DM density profile is
given by
\begin{equation}
\rho(r) = \frac{\delta_c \rho_{\rm c}^0}{(r/r_{\rm s})(1+ r/r_{\rm s})^2}
\end{equation}
where $r_{\rm s}$ is a scale radius, $\rho_{\rm c}^0$ the present
critical density and $\delta_{\rm c}$ a characteristic over-density.
The latter is defined by $\displaystyle \delta_{\rm c}= \frac{100\, c^3g(c)}{3}$,
where $c = r_{\rm v}/r_{\rm s}$ is the concentration parameter of the
halo, $r_{\rm v}$ its virial radius, and $\displaystyle g(c) = (\ln (1+c) - c/(1+c))^{-1}$. 
The concentration $c$ has been found
to correlate with the halo virial mass $M_{\rm v}$ in the range $10^{11}$-$10^{14} h^{-1}$ \sm 
\citep{nfw1997, bullock2001, wechsler2002}, 
so that at a given redshift more massive haloes have lower concentrations. 
In principle, the relation between $c$ and $M_{\rm v}$ makes the NFW model completely 
defined by one parameter (e.g., $M_{\rm v}$). 
However at a fixed mass the scatter in the predicted concentration is large, of the order of 
$\Delta (\log_{10} c) = 0.18$ \citep{bullock2001},
 thus we do not consider the NFW density
profile as a one-parameter family, but we describe it both by the
concentration $c$ and by the virial mass or the circular velocity at
the virial radius. 

At large radii (for $r \gg r_{\rm s}$), the
density behaves as $\rho\propto r^{-3}$, and therefore, the total mass
diverges logarithmically. The resulting mass profile is
\begin{equation}
M(<r) = M_v \, \frac{f(x)}{f(c)} 
\end{equation} 
where $x=r/r_s$ and $\displaystyle f(x) = \ln (1+x) - \frac{x}{1+x}$ \citep{klypin2002}.

When integrating Eq.~\ref{eq:jeans_binneymamon1982},
we set the upper integration limit to $r_{\rm v}$ where we use $r^{2
\beta} \rho_* \sigma_{r,*}^2|_{r_{\rm v}} = 0$ (we are essentially assuming that the 
particles are bound out to the virial radius). As we will see in the next Section and as 
it is found in previous works on dSphs, 
the extent of 
the luminous matter is typically one order of magnitude smaller
 than the extent of the best-fitting DM halo, thus setting the upper integration 
limit to $r_{\rm v}$ instead of to infinity will not affect our results, only decrease the computation time. 

\subsection{Results}\label{sec:res_mass}
We perform the fitting procedure for the observed l.o.s. profile derived in Sect.~\ref{sec:disp_all}, 
i.e. ``Binning2'' \footnote{We checked 
that using the ``Binning3'' velocity dispersion profiles produces best-fitting parameters and models that are  
very similar to those obtained when adopting the ``Binning2'' profile, although in this case 
the minimum $\chi^2$ values for the best-fitting models are larger than the ones from ``Binning2''.}. 
The parameter $p$ in Eq.~\ref{eq:chisquare} denotes the mass enclosed within the last measured point 
for the isothermal sphere and the virial mass for the NFW model; the parameter $p_{\beta}$ is 
$p_{\beta}= \beta$ for the models using $\beta$ constant with radius and 
$p_{\beta}= r_a$ for the models using $\beta_{\rm OM}$.

\begin{figure*}
\begin{center}
\includegraphics[width=0.8\textwidth]{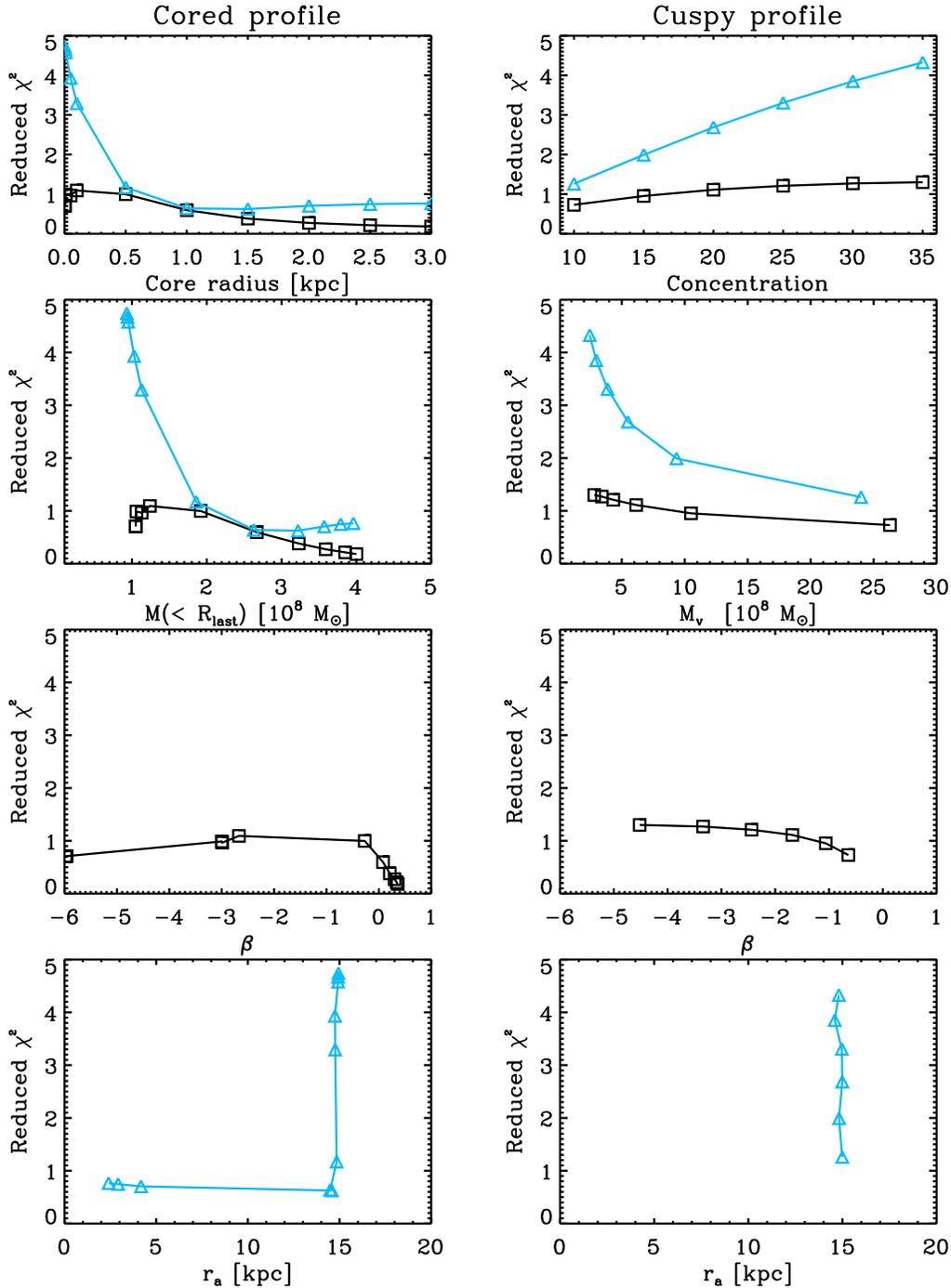}
\caption{In this figure we summarize the results of the mass modelling when 
using the observed l.o.s. velocity dispersion profile derived with ``Binning2'' 
(left: cored dark matter profile; right: cuspy dark matter profile). From top to 
bottom we plot the reduced $\chi^2$ values for the best fit model as a function 
of core radius (left) and concentration (right), 
mass within the last measured point at $\sim$2.3 kpc (left) and virial mass (right) both in units 
of $10^8$ \sm, velocity anisotropy 
in the hypothesis of constant anisotropy with radius, anisotropy radius in the hypothesis 
of velocity anisotropy following an Osipkov-Merritt profile. The black squares and cyan triangles  
show the results in the hypothesis of $\beta = const$ and $\beta = \beta_{\rm OM}$, respectively. 
For each of the core radius (concentration) values shown in the top left (right) panel, 
the models are optimized over the anisotropy and mass paramaters. The quoted reduced $chi^2$ values  
are therefore the value of the $\chi^2$ for the best fit model over 4 degrees of freedom.
}
\label{fig:chisq_summary2}
\end{center}
\end{figure*}

\begin{figure*}
\begin{center}
\includegraphics[width=0.45\textwidth]{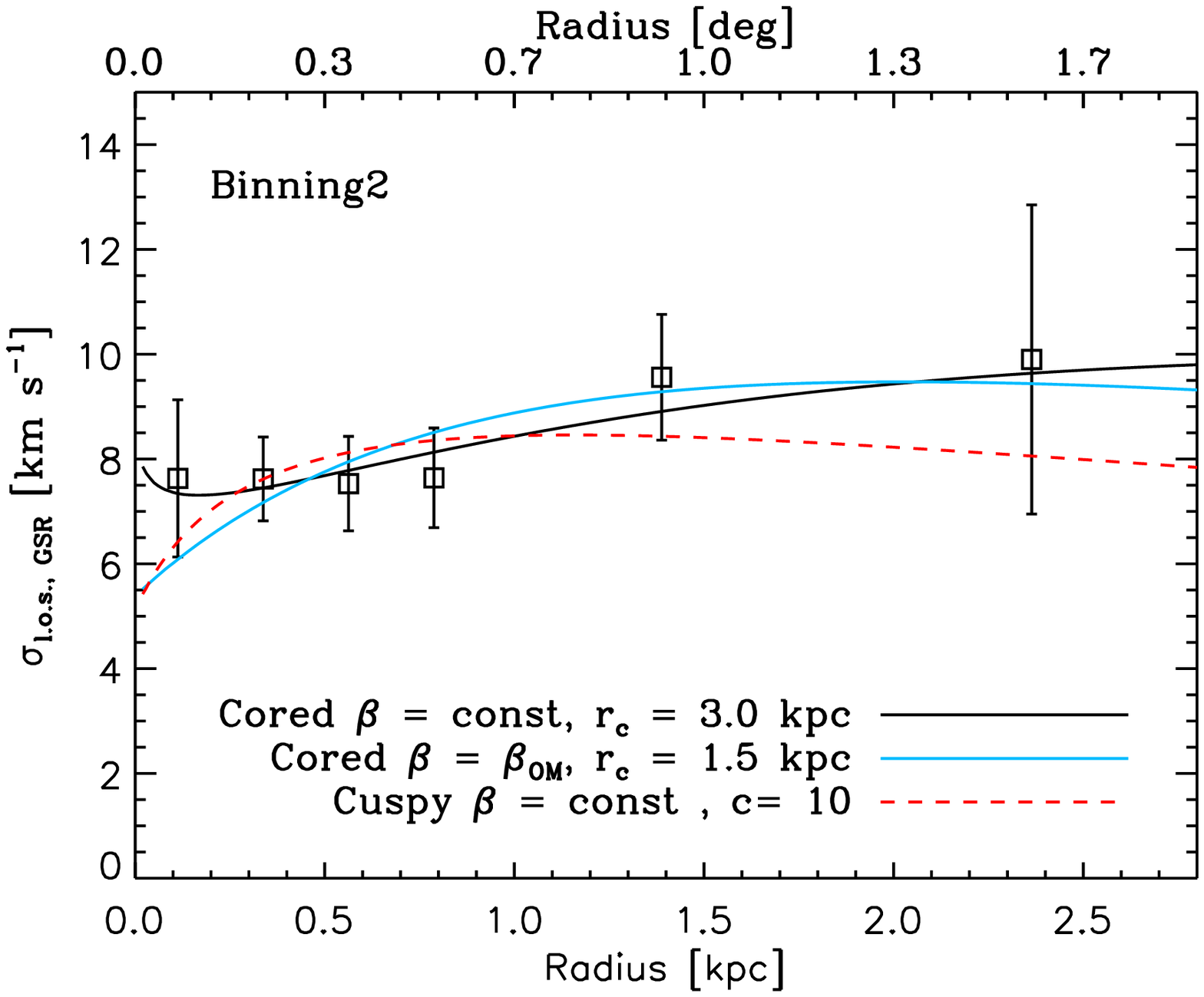}
\caption{Observed l.o.s. velocity dispersion profile (squares with error-bars) overlaid on the best fit model 
for a cored profile with $\beta$ constant (solid black line), a cored profile with $\beta_{\rm OM}$ 
(solid cyan line), cuspy profile with $\beta$ constant (dashed line) for ``Binning2''.  
}
\label{fig:disp_bestfit}
\end{center}
\end{figure*}

The results are summarized in Tab.~\ref{tab:parfit2}.
\begin{itemize}
\item {\bf Isothermal sphere:} In this model the total mass of the DM halo is not finite, therefore we choose to fit 
the value of the mass within the last measured point, $M(< R_{\rm last})$, 
which for ``Binning2'' is $R_{\rm last}=$2.36 kpc. 
 The other free parameter to take into account is the velocity anisotropy $\beta$. 

We explore the performance of models 
with core radii $r_c=$ 0.001, 0.05, 0.1, 0.5, 1, 1.5, 2, 2.5 and 3 kpc. 
In practise, we fix the core radius to each of these values and 
minimize the $\chi^2$ to obtain the best-fitting $M(< R_{\rm last})$ and $\beta$. The results of the fits are summarized in the plots shown in Fig.~\ref{fig:chisq_summary2} 
(left). At a given core radius, the best fit models with constant $\beta$ give in general smaller 
$\chi^2$ values than when $\beta = \beta_{\rm OM}$. The 
models for $\beta = \beta_{\rm OM}$ and $r_c < 0.5$ kpc have a reduced $\chi^2 > 3$, 
which for 4 degrees of freedom corresponds to a significance level of less than 2\%; 
they can therefore be considered as statistically unacceptable. 
All the other best fitting models for the explored core radii have reduced $\chi^2 \la 1$, 
therefore they are all statistically acceptable.  
For these models the best-fitting $M(< R_{\rm last})$ ranges from $\sim$1 - 4 $\times 10^8$ \sm 
(smaller core radii, smaller masses); 
the best-fitting anisotropies ranges from $-6$ to 0.5 for $\beta = const$ (smaller core radii 
more tangential anisotropies), 
while the anisotropy radius goes from 2 to 15 kpc (the upper limit of the explored $r_a$), forcing the anisotropy to be 
close to zero. The models with the smallest reduced $\chi^2$ (see Fig.\ref{fig:disp_bestfit}) 
for $\beta = const$ have $r_c = 3$ kpc (M($<  R_{\rm last}$) $= 4 \pm 0.8 \times 10^8$ \sm and $0.06 < \beta < 0.6$) 
and for $\beta = \beta_{\rm OM}$ $r_c = 1.5$ kpc (M($<  R_{\rm last}$) $= 3.2 \pm 0.7 \times 10^8$ \sm and $r_a > $1.5 kpc). 

This tendency for increasing tangential anisotropy with decreasing core radius comes from the fact that 
smaller core radii models have higher concentrations of mass in the central regions and this tends to 
increase the central value of the velocity dispersion. To match a nearly constant velocity dispersion profile 
the anisotropy then needs to be tangential.

\item {\bf NFW model:} For this model we let the concentration vary from
 $c=$10, 15, 20, 25, 30, 35, and for each of these values we derive the best-fitting virial mass 
and velocity anisotropy. Using an extrapolation of the formulae in the N-body simulations of \citet{jing2000} 
\citep[see][]{koch2007leo1}, 
the concentrations adopted here would correspond to 
the concentrations expected for haloes with virial masses in the range $10^6$-$10^{12}$ \sm 
(smaller mass larger $c$) well containing the range of virial masses expected for dSphs.  

At fixed concentration, all the models with $\beta = \beta_{\rm OM}$ have reduced  $\chi^2$ 
larger than then models with $\beta = const$. The models with c $\ge$ 20 have a reduced $\chi^2 > 2.5$, 
corresponding to a significance level of less than 5\% for 4 degrees of freedom 
and have therefore low statistical significance. The best result for the $\beta = const$ case 
is also given by the model with $c=10$, with a virial mass $M_{\rm v} = 2.6 \pm 0.8 \times 10^9$ \sm and 
$\beta = -0.6_{-1.4}^{+0.6}$ (see Fig.\ref{fig:disp_bestfit}).  
All the models with $\beta = const$ have reduced $\chi^2$ between 1 and 1.5, 
where the virial masses range from approximately 
3$\times 10^8$ to 2.6$\times 10^9$ \sm (smaller $c$ larger virial mass), which gives 
a mass within the last measured point of 1-2$\times 10^8$ \sm, and 
$\beta$ goes from about $-4.5$ to $-0.6$ (smaller $c$ less tangential $\beta$). 
This also 
explains why most of the models with $\beta = \beta_{\rm OM}$ are not favoured for NFW profiles: 
in the hypothesis of $\beta = const$, which yields a good fit to the data, the 
best-fitting anisotropies are mildly tangential; since in the $\beta = \beta_{\rm OM}$ 
the anisotropy is forced to be positive, than the resulting model cannot 
reproduce the data as well. 
\end{itemize}

We conclude that because of the well-known mass-anisotropy degeneracy, it is not 
possible to firmly distinguish between a cored and a cuspy dark matter distribution, nor 
about different hypothesis of anisotropy, solely on the basis of the line-of-sight 
velocity dispersion profile. The only models that can be excluded among the explored ones 
because of their large $\chi^2$ values 
are the NFW models with $\beta = \beta_{\rm OM}$ and $c \ge 20$. However, the general trend is for the 
best-fitting models to have $\beta$ close to isotropic, be it slightly tangential as for the 
NFW case or slightly radial as for the cored profile. Also, the tendency for cuspy profiles of 
low concentrations to produce better fits than for larger concentrations seems to suggest that 
shallower profiles are preferred. Independently on the model, we find that the mass within the last measured point 
of Sextans, at 2.3 kpc, should be in the range 2-4$\times 10^8$ \sm. This would give a 
dynamical mass-to-light ratio between 460 and 920 (M/L)$_{V,\odot}$.

The presence of the central substructure does not affect the mass estimate derived. 
Although the velocity dispersion for stars at the distance covered by the substructure, 
7.6$\pm$1.1 \kms, is lower than if these stars had not been included, i.e. 8.6$\pm$1.4 \kms, 
these two values are consistent within the errors. 
 
\section{Discussion} \label{sec:disc}
$\bullet$ {\bf Metallicity properties:} The stars in Sextans show a wide [Fe/H] range, with the 
majority having [Fe/H] values between $-3.2$ and $-1.4$, with an average [Fe/H] $= -1.9$. 
Among the sample of probable members, 11 extremely metal poor stars are found, reaching down to  [Fe/H] $= -3.8$. If the 
stars in Sextans were formed from a pre-enriched medium, then the presence 
of stars with [Fe/H] as low as $-3.8$ suggests that this interstellar medium 
most likely had not been pre-enriched to a larger value. 

Sextans exhibits a clear spatial variation of its metallicity properties:  
the region at projected radius $R<$ 0.8 deg displays the whole range of metallicities, while at larger 
projected radii practically only stars more metal poor than [Fe/H]$\sim -2.2$ are present. 
Photometric studies of its stellar population show an age difference of 
at least 3 Gyr between the central and outer parts of Sextans, with the central parts 
containing stars between 10 and 14 Gyr old and the outer parts mostly 14 Gyr old stars. It appears 
then that the outer parts were formed very quickly, reaching a metallicity up to [Fe/H]$-2.2$, 
while the inner parts could continue their evolution for at least 3 more 
Gyr and enrich to [Fe/H]$\sim -1.4$. One can speculate that the less prolongued star formation 
of the outer parts may have been caused either from external processes, 
which removed the gas from the external region, or to internal effects, which depleted the outer parts of gas. 
It is although possible that gas was still present in the outer regions for some time 
but simply that conditions for star formation were not met. 

Similar, although less pronounced, spatial variations of metallicity properties have been 
detected in other dSphs, such as for example in Sculptor \citep{tolstoy2004}, 
Fornax \citep{battaglia2006} and to a 
minor extent in Carina \citep{koch2006}. There are hints that Leo~I may exhibit  
such variations \citep{gullieuszik2009} although the current data are 
not sufficient to put this result on a firm ground. Using Str\"omgren photometry 
\citet{faria2007} show the presence of a possible spatial variation in the metallicity 
properties of the Draco dSph. It is still unclear what is the driving process behind these  
metallicity variations. To disentangle environmental effects from internal evolution it is 
important to obtain observations of the stellar populations, kinematics and metallicity 
variations of isolated dwarf galaxies, outside the halo of large spirals. 

$\bullet$ {\bf Velocity gradient:} A statistically significant velocity gradient is found 
along the projected major axis and an axis at P.A.$=191^{\circ}$ in our sample of 
174 probable members of Sextans. \citet{walker2008} also detect a velocity gradient 
in Sextans, although with lower amplitude and approximately along the minor axis of the galaxy. 
Pinning down the amplitude and direction of such gradients, expected to be small for dSphs, 
would require large sample of stars not only in the central regions of the object but also 
in the outer parts, so that the trends along various directions may be accurately determined. 
Unfortunately this is very challenging for a system like Sextans, because of its large extent on the sky and the 
outer parts being sparsely populated. 

Also a univocal interpretation of this feature does not appear possible at this stage. One possibility is that the gradient may 
have originated as the result of tidal disruption of Sextans from the Milky Way, as predicted in N-body models 
of disrupted satellites 
\citep[e.g.][]{oh1995}. However no clear signs of tidal disruption such as 
tidal tails or S-shaped contours are found in Sextans that could confirm this hypothesis, although 
this may be just due to the observational difficulties in detecting such features 
\citep[e.g.][]{munoz2008}, especially in a diffuse and 
heavily contaminated object like Sextans. Knowledge of the accurate proper motion of the galaxy would allow 
to reconstruct its orbital history and determine if strong interactions with the Milky Way 
may have occurred. However no direct determinations of the proper motion of Sextans are available yet in the literature. 
Such determinations would also allow to understand whether the observed gradient may merely be 
a geometrical effect due to the projection of the object transverse motion 
along the line-of-sight because of its large extent on the sky, since in this case the direction of the maximum gradient is expected to be aligned 
with the proper motion direction. 

Another possible option is that the velocity gradient is due to intrinsic rotation of the object. In this case, 
if the flattening of Sextans is due to intrinsic rotation, then it  
would be reasonable to expect the maximum gradient along the projected major axis of the object, as detected in our data. 
When rotation is responsible for the flattening of an axisymmetric galaxy, 
a simple relation holds between the true ellipticity of the galaxy and its mass weighted rotational velocity over the mass weighted 
velocity dispersion, $v_0/\sigma_0$ \citep{binney1987}. As for estimate of the rotational velocity 
we use the fitted value of the gradient at $R \sim 0.7$ deg, the distance of the outermost stars along the major axis which 
cause the rotation signal in our data-set: this value is $v_{\rm rot, los}= 5.9$ \kms and it is likely to represent a lower limit to the 
maximum rotational velocity. Replacing $\sigma_0$ by the velocity dispersion measured along the line-of-sight, 
and $v_0$ by $v_{\rm rot, los}$, the ratio $v_{\rm rot, los}/\sigma_{\rm GSR} = 0.7$ is very close 
to the value expected for an ``isotropic rotator'' (0.727) of ellipticity $e=$0.35 \citep[as tabulated in the]{bm1998}. 
This means that the shape of Sextans is consistent with being flattened by rotation, although 
due to the fact that Sextans velocity ellipsoid and inclination are unknown, 
it is possible that velocity anisotropy is also contributing to the flattening of the galaxy. 

At the moment it is still difficult to build a coherent picture not only on the 
cause of velocity gradients in Milky Way dSphs, but of even their presence. 
No significant gradient is found in Leo~II \citep{koch2007leo2} where the coverage is very homogeneous.
\citet{battaglia2008b} found a significant velocity gradient along the projected major 
axis of the Sculptor dSph, confirmed by \citet{walker2009}. 
Gradients of smaller amplitudes have also been detected in Carina and in Fornax \citep{walker2009}, 
in this case along directions close to the projected minor axis of the object. For those two 
objects the amplitude and direction of the gradient are consistent with being due to the 
transverse motion of the object along the line-of-sight, given their measured proper motion. 
However, while Carina has a very homogeneous coverage, this is not the case for Fornax, 
where the outer parts still await careful scrutiny. The outer parts of these objects appear 
to be particularly important in the measurement of velocity gradients, as shown by the case of 
Leo~I: \citet{mateo2008} found a highly statistically significant gradient when considering only 
the stars located at projected distances larger than 400$''$, 
while they found no significant gradient neither when using the sample of stars at smaller projected distances, 
 nor when considering the whole sample (which is dominated in number by the stars at $R < 400''$). Also combining 
data-sets together may give different results: when using only their GMOS dataset 
\citet{koch2007leo1} find no statistically significant gradient in Leo~I, while when combining their GMOS 
dataset to their DEIMOS one the statistical significance of the detected gradient increases.

This wealth of results on the different Milky Way dSphs shows that a coherent picture of 
velocity gradients as intrinsic feature in dSphs is still lacking, and so probably are 
datasets with both large enough statistics and homogeneous coverage to pin down values and directions of these gradients. 

Velocity gradients have 
been detected along the projected major axis of the 2 isolated dSphs of the Local Group: 
Cetus and Tucana \citep[][respectively]{lewis2007, fraternali2009}\footnote{Note though 
that in these two cases directions other than the major 
axis have not been well explored.}. At the distance of Cetus and Tucana, unless 
of invoking three-body interactions or such \citep{sales2007}, tidal stripping from either the 
Milky Way or M31 cannot have been important, therefore it is very unlikely that the detected gradients 
may be attributed to tides. Also, gradients due to the transverse motion of the system are expected 
to be negligible due to the small extent on the sky of these objects. 
This would point to velocity gradients as being an intrinsic feature of these 
systems. 

Also other spheroidal systems in the Local Group do show velocity gradients along their projected major axis. \citet{geha2010} 
found significant rotation in the NGC~147 and NGC~185 dEs. The authors 
calculate that the 
flattening of these galaxies is consistent with the amount of rotation found, 
although part of it may be due to anisotropy. These galaxies are dEs, not dSphs, with 
luminosities of the order of 10$^8$ \sL, about 2 orders of magnitude larger 
than the common dSph, and 1 order of magnitude larger than Fornax. However, 
as discussed in \citet{kormendy2009} and \citet{tolstoy2009}, the dEs in the Local Group (NGC~147, 
NGC~185 and NGC~205) share similarities with the dSphs, suggesting that these may be the high luminosity 
end of the sequence.

$\bullet$ {\bf Mass determination:} We have performed a Jeans analysis of the observed 
line-of-sight velocity dispersion profile of Sextans and find that 
for the best fitting cored profiles the mass within the last measured point ranges 
from 3.2-4$\times 10^8$ \sm, while the mass within the last measured point for the best fitting 
NFW profile is smaller, $1.9 \pm 0.6 \times 10^8$ \sm but consistent within 1$\sigma$ with the determination 
from the cored profiles. This mass range implies 
that Sextans contains between about 450-900 times more dark than luminous matter, in the 
typical range of Milky Way dSphs. 

Other authors have estimated the mass of Sextans at smaller distances from the 
centre. The mass that \citet{walker2007} find within their last measured point at 1.1 kpc 
is 5.4$\times 10^7$ \sm for a constant anisotropy and an NFW halo 
(with best-fitting $M_{\rm v} = 3\times 10^8$ \sm and $c=20$ derived from the 
concentration-virial mass relation in \citealt{jing2000} and best-fitting $\beta = -2$). 
This is consistent with our model 
with c$=20$, which yields $M_{\rm v} = 6_{-1.5}^{+2.5} \times 10^8$ \sm and $\beta= -1.7$ and 
M($<$1.1 kpc)$= 5.5_{-1.4}^{+2.3} \times 10^7$ \sm. Note that in their analysis \citet{walker2007} choose to fix the concentration to the virial mass according 
to the formula of \citet{jing2000} instead of exploring various concentrations for a given virial 
mass like in this work. It is possible that also \citet{walker2007} data would have yielded a 
lower best-fitting concentration, if the correlation $M_{\rm v}$-$c$ was not assumed.

It has been recently shown that the determination of the dark matter mass within a certain distance is robust 
to model assumptions \citep[e.g.][]{strigari2007}. An interesting result has been that the mass enclosed within 0.3 (0.6 kpc) 
spans a narrow range of values \citep{gilmore2007, strigari2007, strigari2008} and 
this could imply a minimum mass scale for the existence of luminous satellites \citep[see also][for a similar suggestion]{mateo1998}. 
In the determination of \citet{strigari2007} the mass enclosed within 0.6 kpc for Sextans 
is M($<$0.6 kpc)$= 0.9_{-0.3}^{+0.4} \times 10^7$ \sm for a dark matter halo consistent with CDM predictions. 
For comparison, our best-fitting NFW model gives M($<$0.6 kpc)$= 2.0 \pm 0.6 \times 10^7$ \sm, consistent 
within 2$\sigma$. Our best-fitting cored profiles instead give M($<$0.6 kpc)$= 0.9 \pm 0.2 \times 10^7$ \sm 
(constant $\beta$) and M($<$0.6 kpc)$= 1.0 \pm 0.2 \times 10^7$ \sm, both 
very similar to the estimate of \citet{strigari2007}. Therefore in this case Sextans would fit in the 
picture of a minimum mass scale for dSphs. Note that the mass within such a small 
distance from the centre, while it appears to come out naturally from CDM models, 
is not very informative on the total dark matter content of the satellite: 
while the enclosed mass within 0.6 kpc spans 1 order of magnitude ($\sim 10^7$ and $\sim 10^8$ \sm), the total 
dark matter mass may span about 3 orders of magnitude, as a consequence of the lack of a tight correlation 
between concentration and virial mass, and of the particular (accretion and orbital) history of the satellite 
\citep{li2009}. 

\citet{walker2009uni, walker2010err} have expanded the above results to the mass enclosed within the 
half-light radius and also propose 
that all dSphs may be embedded in a dark matter halo of similar mass (and mass profile). 
The mass within the half-light radius (estimate from \citealt{walker2010err}, adjusted for the distance to Sextans here assumed) that we derive  
for the best-fitting NFW model (c$=10$) is M($<$0.7 kpc)$= 2.6 \pm 0.8 \times 10^7$ \sm, in remarkable agreement 
with the determination of \citet{walker2010err}\footnote{Also our other NFW models though give a 
very similar enclosed mass (M($<$0.7 kpc)$= 3.0 \times 10^7$ \sm for c$=20$ and M($<$0.7 kpc)$= 3.2 \times 10^7$ \sm for c$=35$), 
consequence of the degeneracy between concentration and virial mass.}.  The best-fitting cored dark matter 
models however give lower values, $1.4 \pm 0.2 \times 10^7$ \sm  ($r_c= 3$ kpc and $\beta = const$) and 
$1.7 \pm 0.4 \times 10^7$ \sm ($r_c= 1.5$ kpc and $\beta = \beta_{\rm OM}$). 
These estimates make Sextans fall below the ``universal'' mass relation for dSph galaxies 
in \citet{walker2010err}. Until degeneracies between different dark matter models are relieved, 
it is possible that universal relations between dSphs mass and other properties may change according to the adopted models. 

Finally, we would like to point out that for small galactic systems such as dSphs the total mass is likely 
to be an important factor in determining their evolution. Indeed the depth of the potential well 
will influence how much of the gas and metals can be retained and therefore the subsequent 
star formation and chemical enrichement history from the system and how much will be expelled 
because of for example supernovae explosions \citep[e.g.][]{revaz2009}. 
Since outside the last measured point one can only extrapolate mass values, it will not be 
possible to determine total dSph masses with the current methods. However, 
determination as far out as possible are still important 
so as to provide constrains for models that aim at reproducing the detailed observed properties of stars in dSphs. 

\section{Summary} \label{sec:summary}
We obtained VLT/FLAMES intermediate resolution ($R \sim 6500$) spectroscopic observations in the NIR 
CaT region for 1036 distinct targets along the line-of-sight to Sextans, with magnitudes and colours 
broadly consistent with RGB stars. This resulted into 789 stars with S/N and error in velocity such as to produce 
reliable line-of-sight velocities and CaT EWs. Among those, 174 are RGB stars probable members of Sextans with 
line-of-sight velocities accurate to $\pm$2 \kms and CaT [Fe/H] measurements accurate to $\pm$ 0.2 dex. 

The vast majority of the Galactic contaminants could be weeded out from the sample using a simple 3-$\sigma$ kinematic cut. We 
refined our membership criteria by using the \mgi line at 8806.8 \AA\, as an empirical indicator of 
stellar surface stellar gravity, and therefore distinguish between 
probable members of the dSph (RGB stars) and Galactic contaminants (most likely dwarf stars). 

We used the sample of probable members to investigate the wide-field metallicity and kinematic properties of Sextans, and to 
perform a determination of its mass content. 

Sextans is a metal poor system, with an average [Fe/H]$=-1.9$. Its stars 
display a wide range of metallicities, with the majority being between $-3.2 <$ [Fe/H] $< -1.4$. 
A spatial variation of the metallicity properties is present, with the central parts 
being in average more metal rich than the outer parts. There are indications of a link 
with the kinematics, with the stars more metal rich than [Fe/H]$=-2.2$ having 
a lower dispersion than the stars with [Fe/H]$ < -2.4$.

The analysis of the radial trend of the average velocity and dispersion of Sextans stars as 
function of metallicity has uncovered the presence of a cold, metal-poor ([Fe/H]$\sim -2.6$ from 
CaT measurements) kinematic substructure at projected 
radii $R < 0.22$ deg. This structure is consistent with being a disrupted stellar cluster. It is possible 
this may be the same feature observed by \citet{kleyna2004}.

Our sample of Sextans stars shows a statistically significant velocity gradient along the projected 
major axis of the galaxy and along an axis at P.A.$=191^{\circ}$ (8.5 $\pm$ 3.0 \kmsdeg and 
7.5$_{-3.0}^{+3.4}$ \kmsdeg, respectively). Given the current global observational understanding of this 
object it is unclear whether this gradient is due to intrinsic rotation, tidal disruption or 
geometrical effects. Also, a better sampling of the outer parts, both in terms of spatial coverage 
and number statistics, would allow to place the amplitude of the velocity gradient on more secure footing  
and perhaps determine the overall velocity pattern throughout the face of the galaxy.

In the hypothesis of dynamical equilibrium, we performed a mass determination of Sextans, 
following a Jeans analysis and comparing the observed line-of-sight velocity dispersion profile to 
the predictions for cored and cuspy dark matter models. Sextans appears to be a very dark matter dominated 
object, with a mass within 2.3 kpc of 2 - 4$\times 10^8$ \sm, giving a 
dynamical mass-to-light ratio between 460-920 (M/L)$_{V,\odot}$. Within the explored 
dark matter models and velocity anisotropy profiles, the best fits are given by cored profiles  
with large core radius and mildly radial anisotropy. However also a NFW profile with 
mildly tangential anisotropy gives a very good representation of the data; the concentration 
of the favoured NFW is $c=10$ somewhat lower for what expected for galaxies in the 
mass range of Sextans when using the virial mass - concentration relation of \citet{jing2000}. 

\section*{Acknowledgements}
G.B. and P.P. are grateful to the Kapteyn Astronomical Institute for hospitality 
and financial support during part of this work. G.B. 
thanks F.Fraternali for useful suggestions. The authors acknowledge E.Starkenburg 
for kindly running preliminary models regarding the \mgi line. Financial support 
by the European Research Council has been provided to A.H. in the form of 
SZG-GALACTICA, \#240271.

\begin{table*}
\begin{center}
\begin{tabular}{lcc}
\hline
\hline
Parameter & value  & reference \\
\hline
($\alpha_{\rm J2000}$,$\delta_{\rm J2000}$) & 10$^h$ 13$^m$ 03$^s$ $-$01$^{\circ}$ 36$'$ 54$''$ & 1 \\
P.A. & 56$^{\circ}$ & 2 \\
$e$  & 0.35 & 2 \\
R$_{\rm core}$ & 16.6 arcmin & 2 \\
R$_{\rm tidal}$ & 160 arcmin & 2 \\
R$_{\rm e}$ & 15.5 arcmin & 2 \\
v$_{\rm sys}$  & 226.0$\pm$0.6 \kms & 3\\
v$_{\rm sys, GSR}$  & 78.4$\pm$0.6 \kms & 3\\
$\sigma$       &   8.4$\pm$0.4 \kms & 3\\
$\sigma_{\rm GSR}$       & 8.4$\pm$0.4 \kms & 3\\
(m-M)$_0$        & 19.67  & 1 \\
Distance  & 86 kpc  & 1\\
L$_V$ & $4.37 \pm 1.69 \times 10^5$ \sL  & 2, 4 \\
V$_{\rm HB}$ & 20.35 & 2 \\
E(B-V)       & 0.0477 & 5 \\
\hline
\end{tabular}
\caption{The various rows are from top to bottom: coordinates of the optical centre; 
position angle, defined as the angle between the north direction and the major axis of the galaxy measured 
counter-clockwise; ellipticity, defined as $e=1 - b/a$; King core and tidal radius; exponential radius; 
systemic velocity in the 
heliocentric and in the GSR system; global velocity dispersion, in the 
heliocentric and in the GSR system; distance modulus and heliocentric distance; luminosity in V-band 
(based on the apparent magnitude measured by \citealt{IH1995} but readjusted for a distance of 86 kpc 
by \citealt{lokas2009}); V magnitude of the 
horizontal branch; reddening. References: 1 $=$ \citet{mateo1998}; 2 $=$ \citet{IH1995}; 3 $=$ this work; 4 $=$ \citet{lokas2009}; 
5 $=$ \citet{schlegel1998}, http://irsa.ipac.caltech.edu/cgi-bin/bgServices/nph-bgExec, average value over 5 degrees. 
} \label{tab:par}
\end{center}
\end{table*}

\begin{table*}
\begin{center}
\begin{tabular}{lcccccc}
\hline
\hline
 Field name  &  RA(J2000) & DEC(J2000) & Date & Exp. time [sec] & S/N & N$_{\rm targ}$ \\
\hline
Sext\_100 &     10$^h$12$^m$58.1$^s$ & $-01^{\circ}38'05.5''$ & 2003-12-20 & 2700 &   18.6 &  54 \\
Sext\_101 &     10$^h$14$^m$47.4$^s$ & $-01^{\circ}11'27.5''$ & 2003-12-22 & 3800 & 17.2 &  55   \\
Sext\_102 &     10$^h$13$^m$22.1$^s$ & $-01^{\circ}21'53.6''$ & 2004-03-14 & 500, 1125, 3600 & 2.9,4.5,15.5 & 74,92,92 \\
Sext\_103 &     10$^h$11$^m$13.3$^s$ & $-01^{\circ}12'01.9''$ & 2003-12-22 & 4000 &  13.3 &  50  \\
Sext\_104 &     10$^h$11$^m$21.6$^s$ & $-01^{\circ}31'38.0''$ & 2004-03-15 & 3600 & 21.1  & 68  \\
Sext\_112 &     10$^h$13$^m$38.5$^s$ & $-02^{\circ}00'35.6''$ & 2008-05-25 & 3600 &  14.0 & 81 \\
Sext\_113 &     10$^h$14$^m$52.3$^s$ & $-02^{\circ}24'24.5''$ & 2004-01-01 & 3600 & 25.1 & 48  \\ 
Sext\_114 &     10$^h$12$^m$24.1$^s$ & $-01^{\circ}58'43.5''$ & 2004-03-16 & 4600 & 17.5 & 83  \\ 
Sext\_115 &     10$^h$11$^m$30.4$^s$ & $-02^{\circ}09'45.0''$ & 2003-12-20 & 3600 & 8.8 &  57  \\ 
sext\_ext\_02 & 10$^h$17$^m$03.1$^s$ & $-01^{\circ}01'49.8''$ & 2004-03-17 & 3600 &  26.8 &  76  \\
sext\_ext\_03 & 10$^h$18$^m$36.3$^s$ & $-00^{\circ}24'21.9''$ & 2004-03-18 & 3600 &  28.7 &  69   \\
sext\_ext\_04 & 10$^h$20$^m$43.2$^s$ & $-00^{\circ}06'51.4''$ & 2004-03-19 & 3600 & 37.4 &  79  \\
sext\_ext\_06 & 10$^h$13$^m$22.9$^s$ & $-00^{\circ}26'51.1''$ & 2004-03-20 & 4200 & 42.0 &  68  \\
sext\_ext\_08 & 10$^h$08$^m$27.4$^s$ & $-00^{\circ}52'01.6''$ & 2004-03-17 & 3600 &  31.6 &  58  \\
sext\_ext\_14 & 10$^h$07$^m$42.2$^s$ & $-03^{\circ}13'27.8''$ & 2004-03-20/21 & 2511, 3600 & 29.9,33.3  & 66,66\\
sext\_ext\_16 & 10$^h$07$^m$29.2$^s$ & $-01^{\circ}40'19.7''$ & 2004-03-21 & 4500 & 27.7 &  57  \\
\hline
\end{tabular}
\caption{Journal of our VLT/FLAMES observations at R$\sim$6500 of Sextans. 
The columns give respectively: the name of the observed field, the coordinates 
of the FLAMES pointing, the UT date of observation, the exposure time, the median signal-to-noise and the number of the observed 
spectra. All these observations were taken in service mode by ESO staff. Field 102 and ext\_14 have repeated exposures 
because the first exposures did not meet the required criteria for the maximum allowed seeing and/or airmass. This repeated exposures 
also allow a check for the internal velocity and EWs errors.
} \label{tab:journal}
\end{center}
\end{table*}

\begin{table*}
\begin{center}
\begin{tabular}{lccccccl}
\hline
\hline
  & $\chi^2$ &  $\chi^2/\nu$ & $p_{\beta}$ & M($< R_{\rm last}$) & M$_{\rm v}$\\
\hline
Cored $r_c=$ 3.0 kpc $\beta = const$ & 0.73 & 0.2 & $0.06 < \beta < 0.6$ & $4.0 \pm 0.7 \times 10^8$ \sm & \\
Cored $r_c=$ 1.5 kpc $\beta = \beta_{\rm OM}$ & 2.5 & 0.6 & $r_a > 1.5 $ kpc & $3.2 \pm 0.7 \times 10^8$ \sm & \\
Cuspy c $=$ 10  $\beta = const$ & 2.9 & 0.7 & $-1.4 < \beta < 0 $ & $1.9 \pm 0.6 \times 10^8$ \sm  & $2.6 \pm 0.8 \times 10^9$ \sm \\
\hline
\end{tabular}
\caption{Parameters of the best-fitting dark matter models for mass modelling of Sextans, 
when using the observed l.o.s. velocity dispersion profile from ``Binning2''. 
The columns show: the $\chi^2$, the reduced  $\chi^2$ (with the number 
of degrees of freedom being $\nu=4$), the parameter defining the anisotropy 
(i.e. $\beta$ itself for the $\beta = const$ case, and the anisotropy radius $r_a$ [kpc] for the 
$\beta = \beta_{\rm OM}$ case), the mass contained within the last measured point (at $\sim$2.3 kpc, 
assuming a distance to Sextans of 86 kpc), and for the cuspy profile the virial mass. 
} \label{tab:parfit2}
\end{center}
\end{table*}

\begin{table*}
\begin{center}
\begin{tabular}{lcccccc}
\hline
\hline
Name  & $\alpha_{\rm J2000}$ &  $\delta_{\rm J2000}$ & S/N & v$_{\rm hel}$ [\kms]& 
$\Sigma W$ [\AA\,] & [Fe/H]\\
\hline
sext\_000 &  10 12 40.27 & $-$1 29  7.7 & 13.0 &  241.51 $\pm$    4.64 &    2.28 $\pm$    0.46 & $-2.34_{-0.31}^{+0.27}$ \\
sext\_001 &  10 12 47.98 & $-$1 29 38.8 & 22.7 &  222.51 $\pm$    1.70 &    3.44 $\pm$    0.26 & $-1.78_{-0.13}^{+0.13}$ \\
sext\_002 &  10 12 39.03 & $-$1 29 59.0 & 34.6 &  235.84 $\pm$    1.22 &    3.81 $\pm$    0.17 & $-1.87_{-0.08}^{+0.08}$ \\
sext\_003 &  10 13  9.94 & $-$1 29 34.9 & 21.7 &  241.29 $\pm$    2.81 &    2.93 $\pm$    0.28 & $-1.95_{-0.15}^{+0.15}$ \\
sext\_004 &  10 13 18.97 & $-$1 26 58.2 & 20.0 &  216.48 $\pm$    2.40 &    1.96 $\pm$    0.30 & $-2.53_{-0.23}^{+0.20}$ \\
sext\_005 &  10 13 17.14 & $-$1 26 38.1 & 22.9 &  245.40 $\pm$    1.48 &    2.54 $\pm$    0.26 & $-2.27_{-0.15}^{+0.14}$ \\
\hline
\end{tabular}
\caption{This table lists the relevant data for the stars in Sextans observed with VLT/FLAMES 
which passed our membership criteria. 
The columns indicates: (1) the star ID; 
(2),(3) star coordinates (right ascension in hours and declination in degrees); (4) S/N/\AA\, ; 
(5) heliocentric velocity and its error; (6) summed CaT equivalent width ($EW_2 + EW_3$) and its error; 
(7) [Fe/H] value and error, derived adopting the calibration from \citet{starkenburg2010}.
} \label{tab:data}
\end{center}
\end{table*}

\newpage

\clearpage

\clearpage

\clearpage

\clearpage

\clearpage

\bibliographystyle{mn2e} 
\bibliography{sextans_biblio}

\section*{Appendix: Comparison with Walker et al. (2009)}
We compare our dataset of objects with S/N$\ge$10 and velocity errors $\le$ 5 \kms to the recent 
data-set of \citet{walker2009} consisting of spectroscopic observations 
of 947 distinct targets along the line-of-sight to Sextans  
acquired with the Michigan/MIKE Fiber System (R$=$20000-25000). These 
spectra cover the wavelength range 5140-5180 \AA\, including the magnesium triplet 
absorption feature which the authors use, together with calibrating globular 
clusters, to provide an estimate of [Fe/H] for the individual stars. 

\subsection*{Velocities}
There are 141 overlapping individual stars with velocity determinations between our and 
\citet{walker2009} samples, including probable members and non-members. 
In Fig.~\ref{fig:comp_walker} (left) we compare the line-of-sight velocities for the 
stars overlapping between the two samples. 

The agreement between the 
velocity determinations in the two data-sets is good: the weighted average velocity difference between the 
two samples is 1.3$\pm$1.3 \kms, with a scaled MAD of 3.8 \kms (corresponding to about an average 
velocity error of 2.5 \kms in each of the two datasets) and 1.4 for the distribution 
of velocity differences normalized by the errors. Overall the agreement is good, except 
for 5 stars, which have velocities consistent with membership to Sextans in our sample but 
heliocentric velocities below 20 \kms in the \citet{walker2009} sample, which would 
classify them as clearly MW stars: the velocities of these 
differ by about 250-300 \kms between the two samples. All these stars have very reasonable S/N in our data and 
visual examination of the 
spectra of these 5 stars with discrepant velocities does not reveal any 
problem with the spectrum (i.e. CaT is in general visible enough to allow a 
reasonable velocity estimation with our routine). 
We find it unlikely that these may all be binary stars, 
but the reason for this discrepancy remains unclear. 

Excluding these 5 stars does not significantly alter the quality of the comparison 
between the two samples, bringing some improvement in the average 
velocity difference between the two samples (0.41$\pm$0.44 \kms), while the 
scaled MAD remains almost unchanged. 

\subsection*{Metallicities}
\begin{figure*}
\begin{center}
\includegraphics[width=70mm]{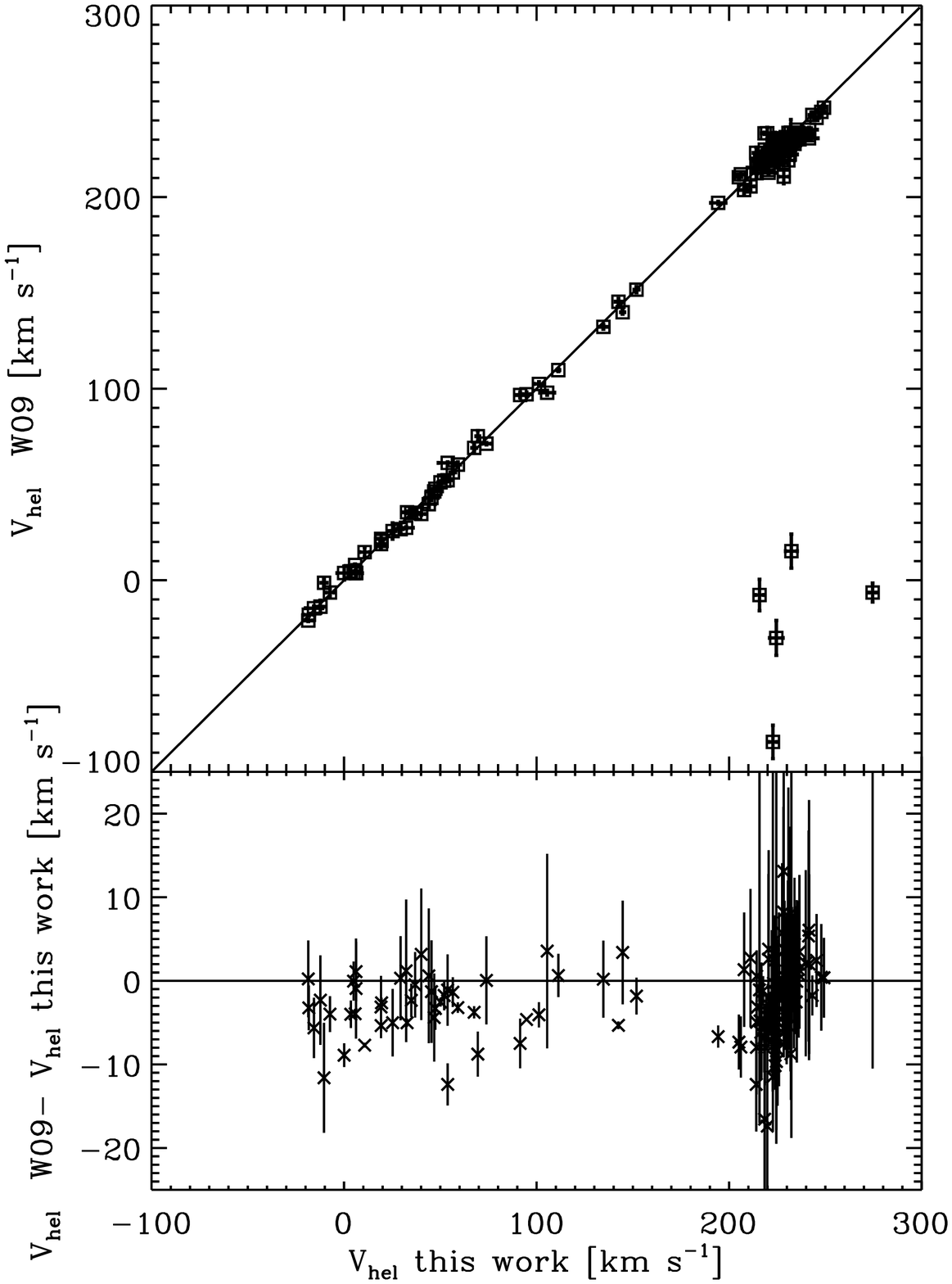}
\includegraphics[width=70mm]{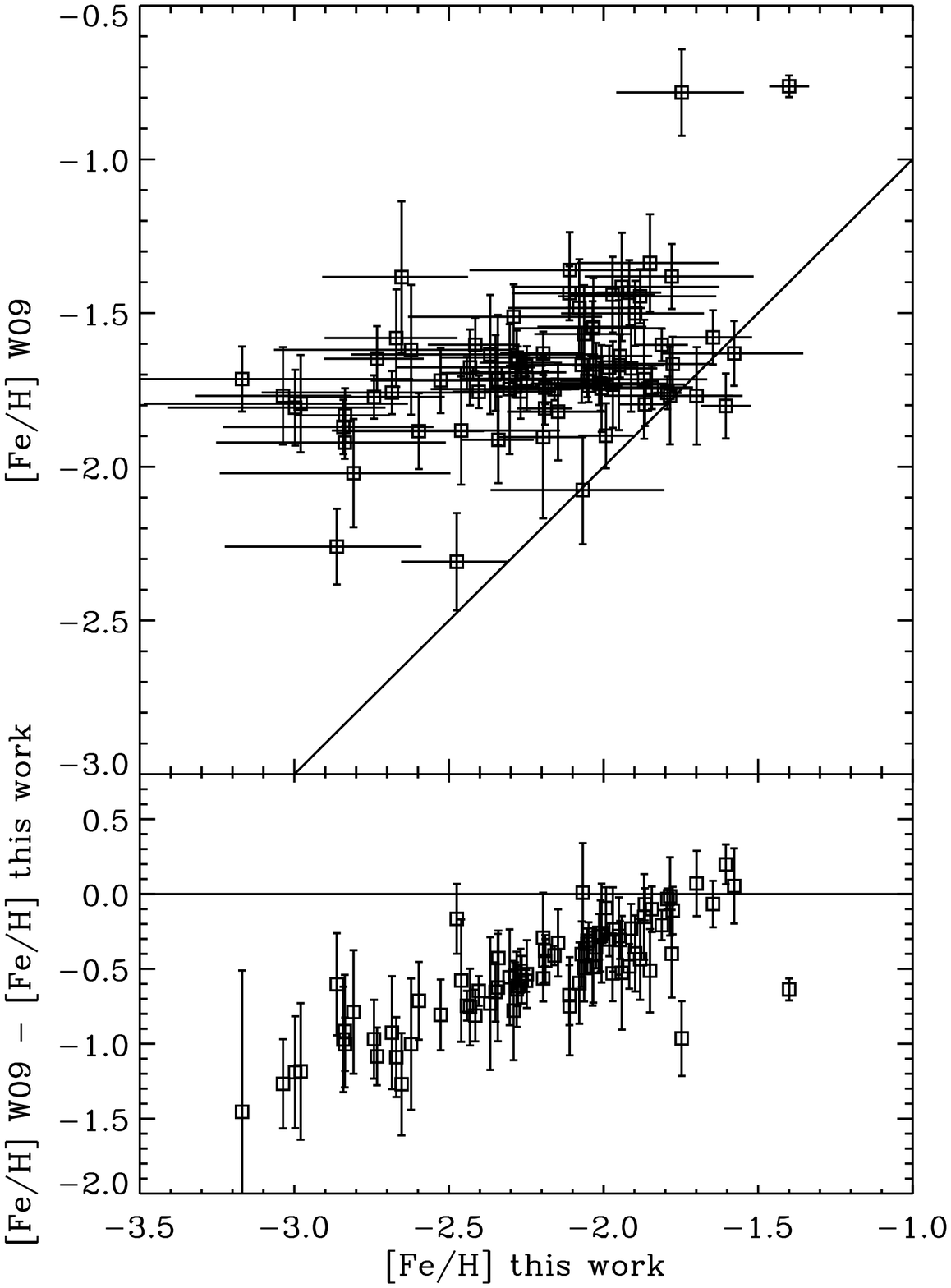}
\caption{Comparison between overlapping stars between this work and \citet{walker2009}. 
The comparison for the velocity determinations is shown in the left panels 
(134 stars, both probable members and non-members), and for the [Fe/H] values 
in the right panels (94 stars, only probable members). The top panels show the direct comparison, with the solid line 
indicating the one-to-one relation; the bottom panels show the differences among the 
determinations as a function of the quantities derived in this work, with the 
solid line indicating a null difference.}
\label{fig:comp_walker}
\end{center}
\end{figure*}

\begin{figure*}
\begin{center}
\includegraphics[width=70mm]{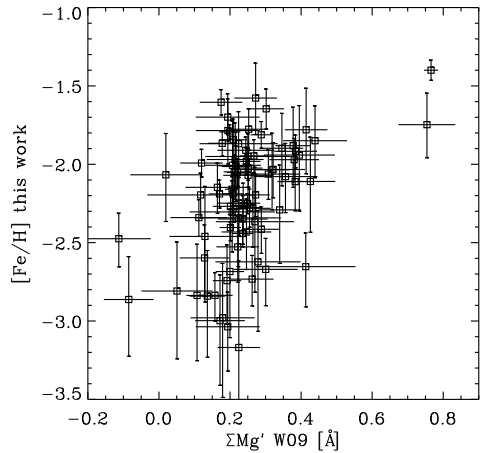}
\caption{[Fe/H] values as determined in this work against determinations of the Mg spectral index from 
\citet{walker2009} for 94 Sextans probable members which overlap between the two samples. 
Note the large range of [Fe/H] values 
corresponding to a narrow range in spectral index values.}
\label{fig:mgp_walker}
\end{center}
\end{figure*}

Since the [Fe/H] calibrations in both studies are based on different spectral features, 
a comparison between the [Fe/H] determinations of the individual stars is expected to be meaningful 
only when considering probable Sextans members, and not for the MW contaminants for which the 
luminosity correction for the dwarf stars is likely to be incorrect and different in the two studies. 
\citet{walker2009} derive [Fe/H] measurements 
from a Mg index and place them on the \citet{carretta1997} scale (their Eq.~7). 
The calibration of the spectral indexes to a [Fe/H] scale requires a correction for the V-magnitude of the 
stars, therefore the authors provided [Fe/H] values only for those Sextans stars in common between their sample 
and the photometric study of \citet{lee2003}, using the V-magnitudes derived in the latter work. 
This would result in 50 probable Sextans members with [Fe/H] estimates in common between the two studies.
 In order to increase the number of stars available for comparison with our sample, 
we check that the V-magnitudes obtained from our photometry compares well to the determinations of \citet{lee2003} for the 
same stars, and then use our V-magnitudes for those stars in common between our sample and Walker's but that did not have 
determinations from \citet{lee2003}: this gives an additional 44 probable members in common between the two samples. 
The results of the comparison are shown in the right panel of Fig.~\ref{fig:comp_walker}: in general the 
determinations from W09 yield higher values of [Fe/H] (more metal rich) and the [Fe/H] distribution appears 
much narrower. This effect was also highlighted by \citet{walker2009} and the authors themselves advise 
against using the determinations from the Mg index as absolute values. The authors 
suggest that the Mg spectral index could be used as a relative indicator of metallicity. 
In Fig.~\ref{fig:comp_walker} (right) we show that the metallicities from the Mg spectral index correspond to 
a narrower range than the more reliable CaT measurements, suggesting a limited use of the Mg index also for 
relative measurements. It appears also not possible to carry out a reliable 
calibration between [Fe/H] estimates from CaT and from Mg index 
because of the large scatter between the two quantities (Fig.~\ref{fig:mgp_walker}).

The fact that the Mg index is not a good indicator of [Fe/H] prevents us from combining our sample of metallicities 
with those of \citet{walker2009}. However, such a merged data-set would be most helpful 
to considerably increase the size and coverage of both works. 

\label{lastpage}

\end{document}